\newif\ifpreprint
\preprinttrue 

\newif\ifSupplementary
\Supplementaryfalse

\ifpreprint
\documentclass{natureprintstyle}
\else
\documentclass{nature}
\fi

\usepackage{epsfig}
\usepackage{color}
\usepackage{bm}
\usepackage{graphicx}
\usepackage{longtable}
\usepackage{amssymb}
\usepackage{tikz}
\usepackage{paralist}
\usepackage{lscape}

\newcommand{\msol}{\hbox{$M_\odot$}}

\newcommand{\lsol}{\hbox{$L_\odot$}}

\newcommand{\lir}{\hbox{$L_{\mathrm{IR}}$}}
\newcommand{\lco}{\hbox{$L^\prime(\mathrm{CO})$}}

\newcommand\phn{\phantom{0}}

\newcommand{\jone}{\hbox{$J_1$}}
\newcommand{\jtwo}{\hbox{$J_2$}}
\newcommand{\jthree}{\hbox{$J_3$}}
\newcommand{\hs}{\hbox{$H_s$}}
\newcommand{\hl}{\hbox{$H_l$}}

\newcommand\farcs{\mbox{$.\!\!^{\prime\prime}$}}%

\newcommand{\apj}{Astrophys. J.}

\newcommand{\pasp}{Publ. Astron. Soc. Pac.}
\newcommand{\apjs}{Astrophys. J. Supp.}
\newcommand{\araa}{Annu. Rev. Astron. Astrophys.}
\newcommand{\mnras}{Mon. Not. R. Astron. Soc.}
\newcommand{\physrep}{Phys. Rep.}
\newcommand{\apjl}{Astrophys. J. Let.}
\newcommand{\aap}{Astron. Astrophys.}
\newcommand{\aj}{Astron. J.}
\newcommand{\nat}{Nature}

\def\figureOneCaption{
  \caption{\textbf{Images of Milky Way Progenitors at redshifts
      $\mathbf{z}$ = 1.2 to 1.3.}  The top panels show the ALMA images of
    the redshifted CO $J$=3--2 emission for each galaxy.   The inset bar
    shows a scale length of 3 arcseconds, and the hashed ellipse shows the
    size of the synthesized ALMA beam of each observation.  {The contours
      denote the emission at 2 times the noise.}    The bottom panels show
    combined Hubble Space Telescope images at 0.78, 1.1, and 1.6~$\mu$m
    (approximately the rest-frame $U$--, $V$--, and $R$--band emission).
    The contours denote ALMA CO(3-2) emission with levels at 2, $2\sqrt
    2$, and 4 times the noise. The inset bar shows a scale length of 1
    arcsecond, which corresponds to a physical scale of 8.3--8.4~kpc at
    these redshifts. \label{fig:co_hst} }
}

\def\figureOne{
    \setcounter{figure}{0}
    \ifpreprint \begin{figure*} \else \begin{figure} \fi
        \centerline{
          \includegraphics[width=0.5\textwidth]{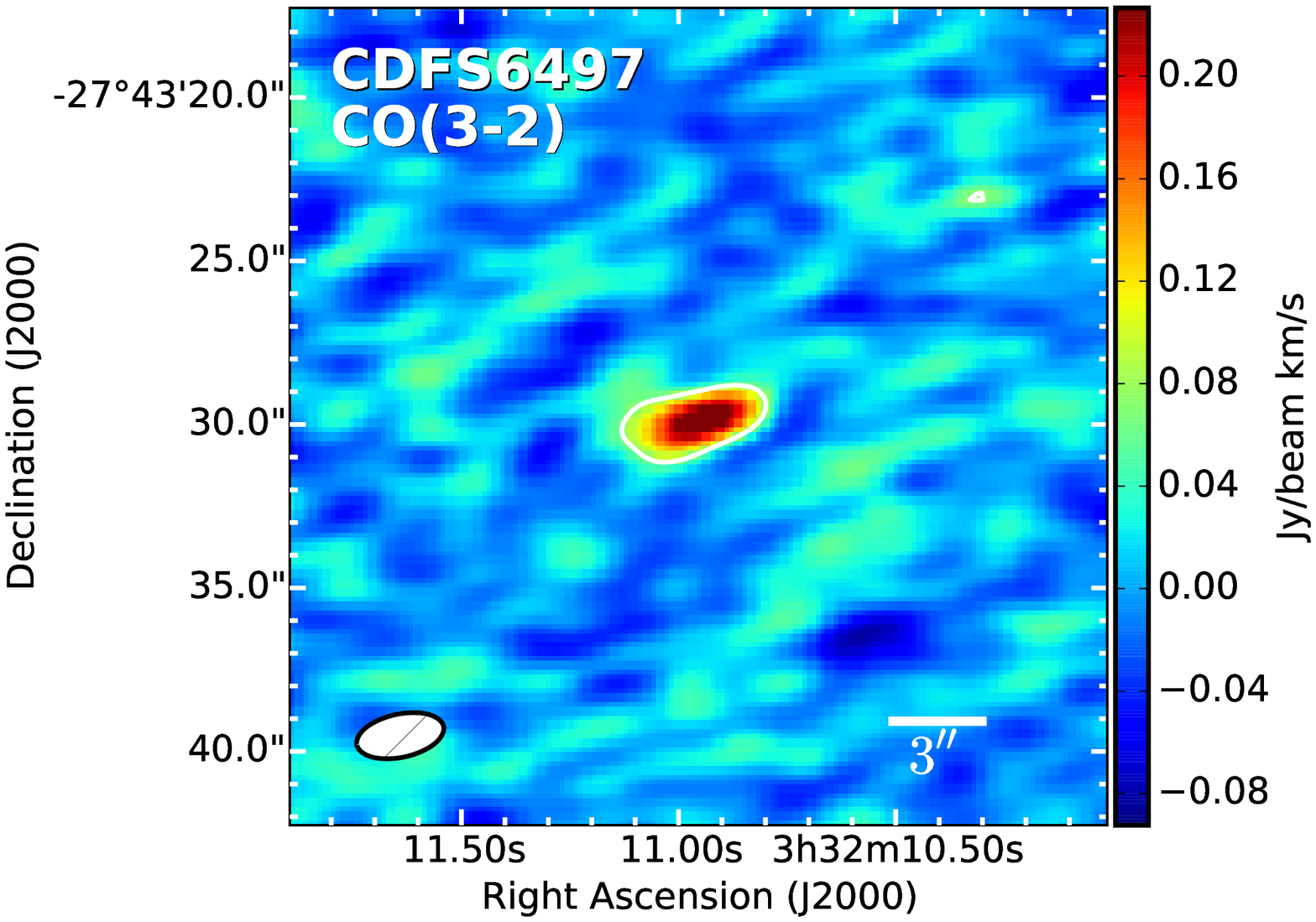}
          \includegraphics[width=0.5\textwidth]{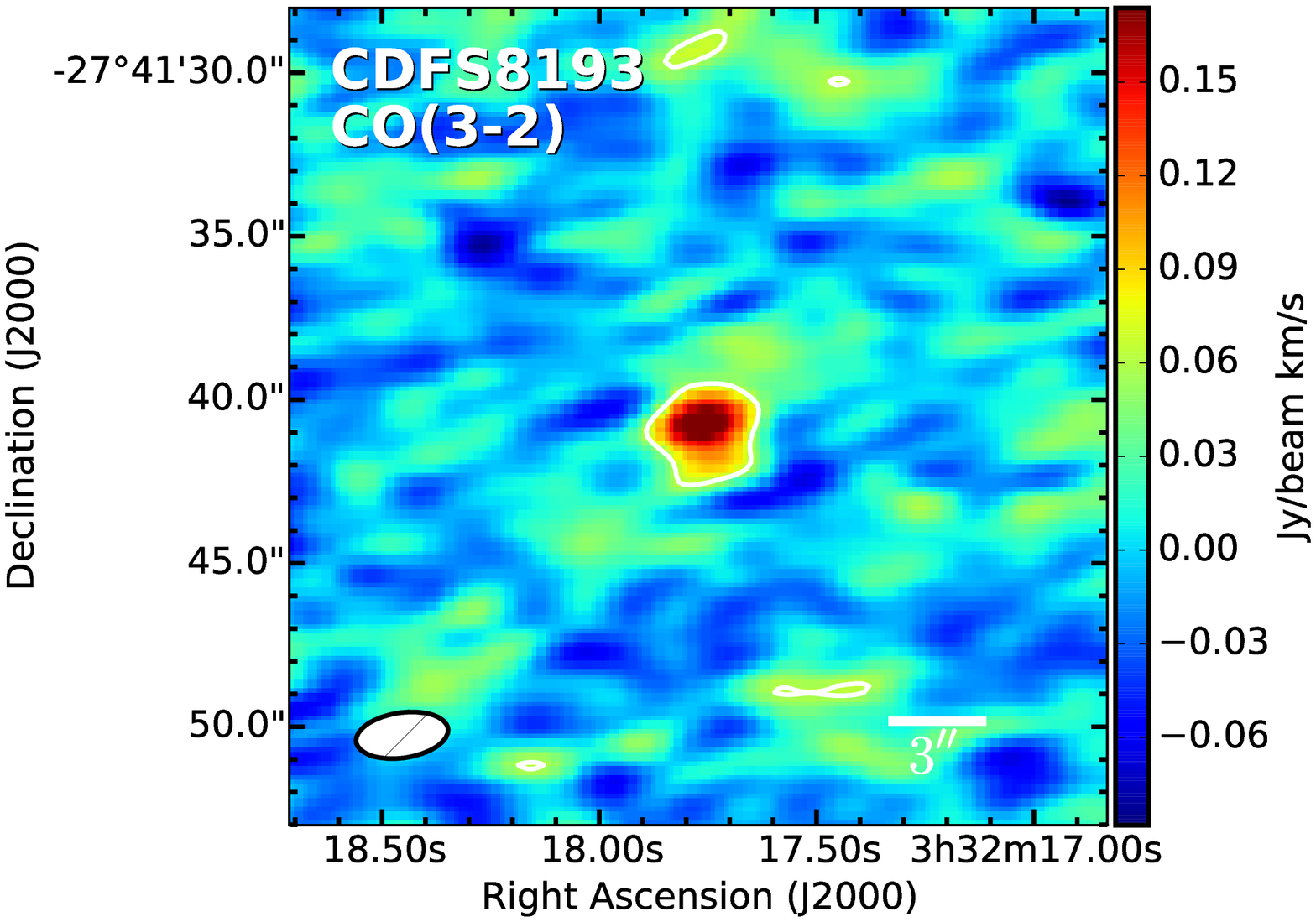}
        }
      \centerline{
        \includegraphics[width=0.5\textwidth]{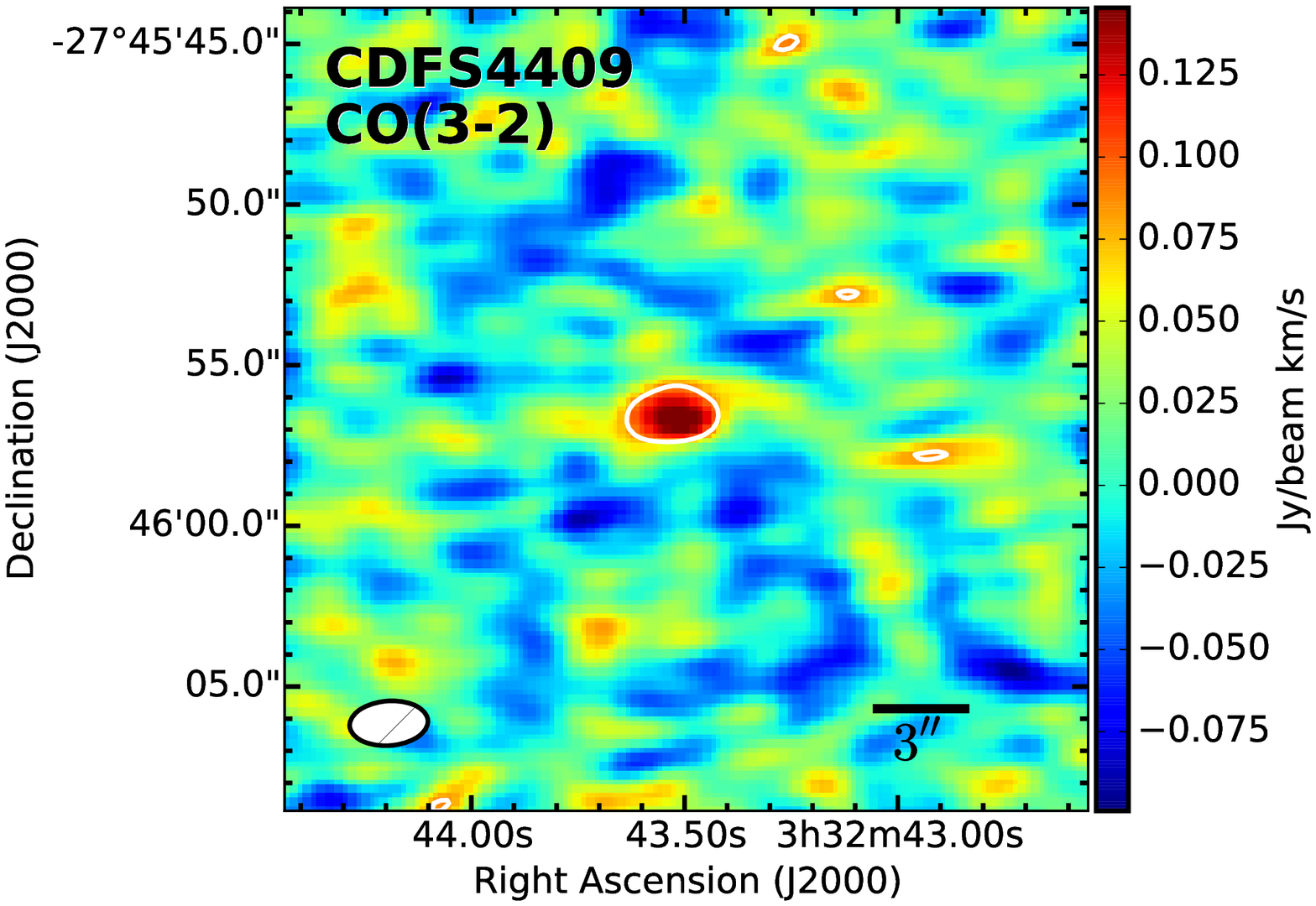}
        \includegraphics[width=0.5\textwidth]{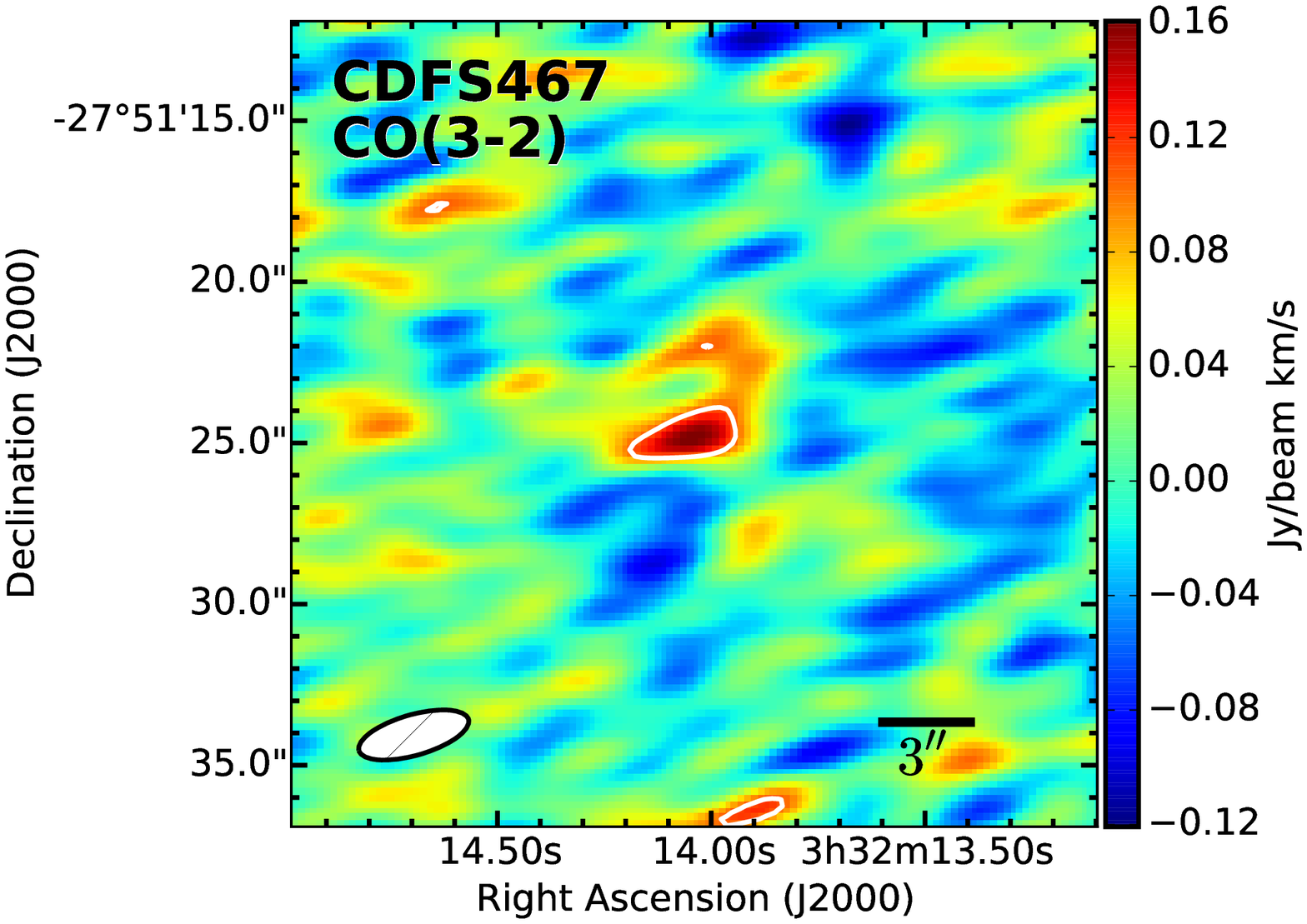}
      }
      \centerline{
        \includegraphics[width=0.25\textwidth]{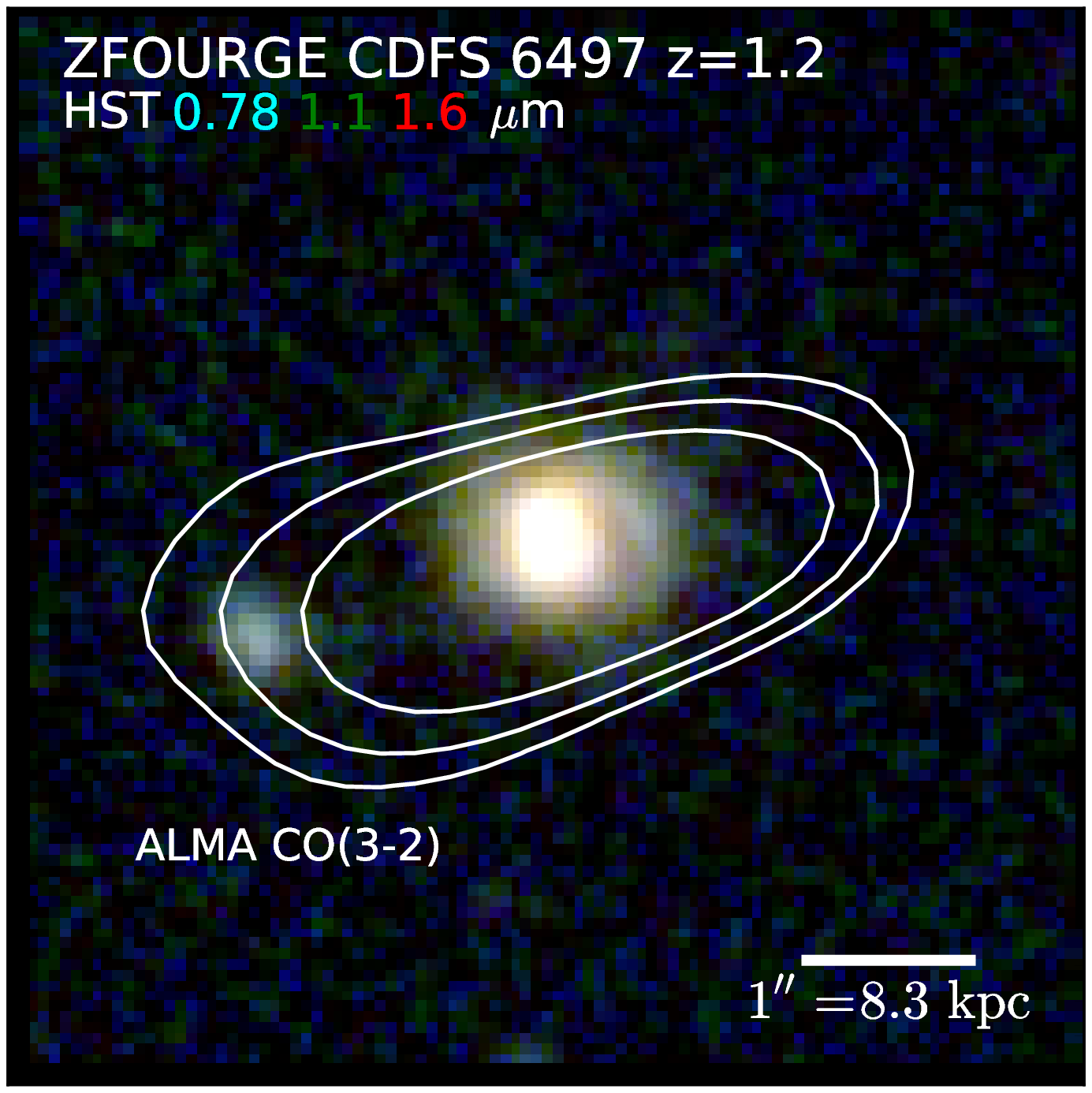}
        \includegraphics[width=0.25\textwidth]{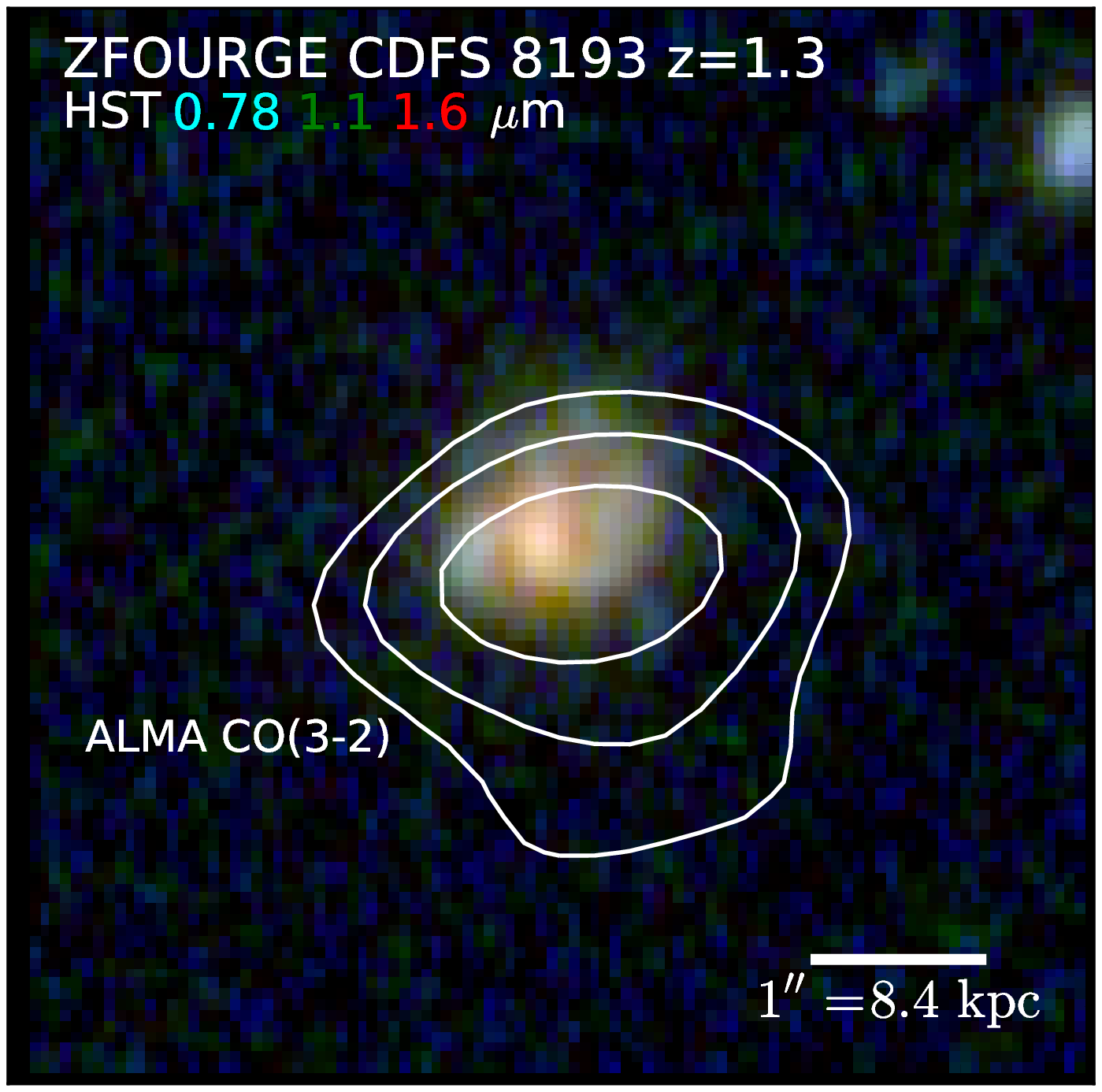}
        \includegraphics[width=0.25\textwidth]{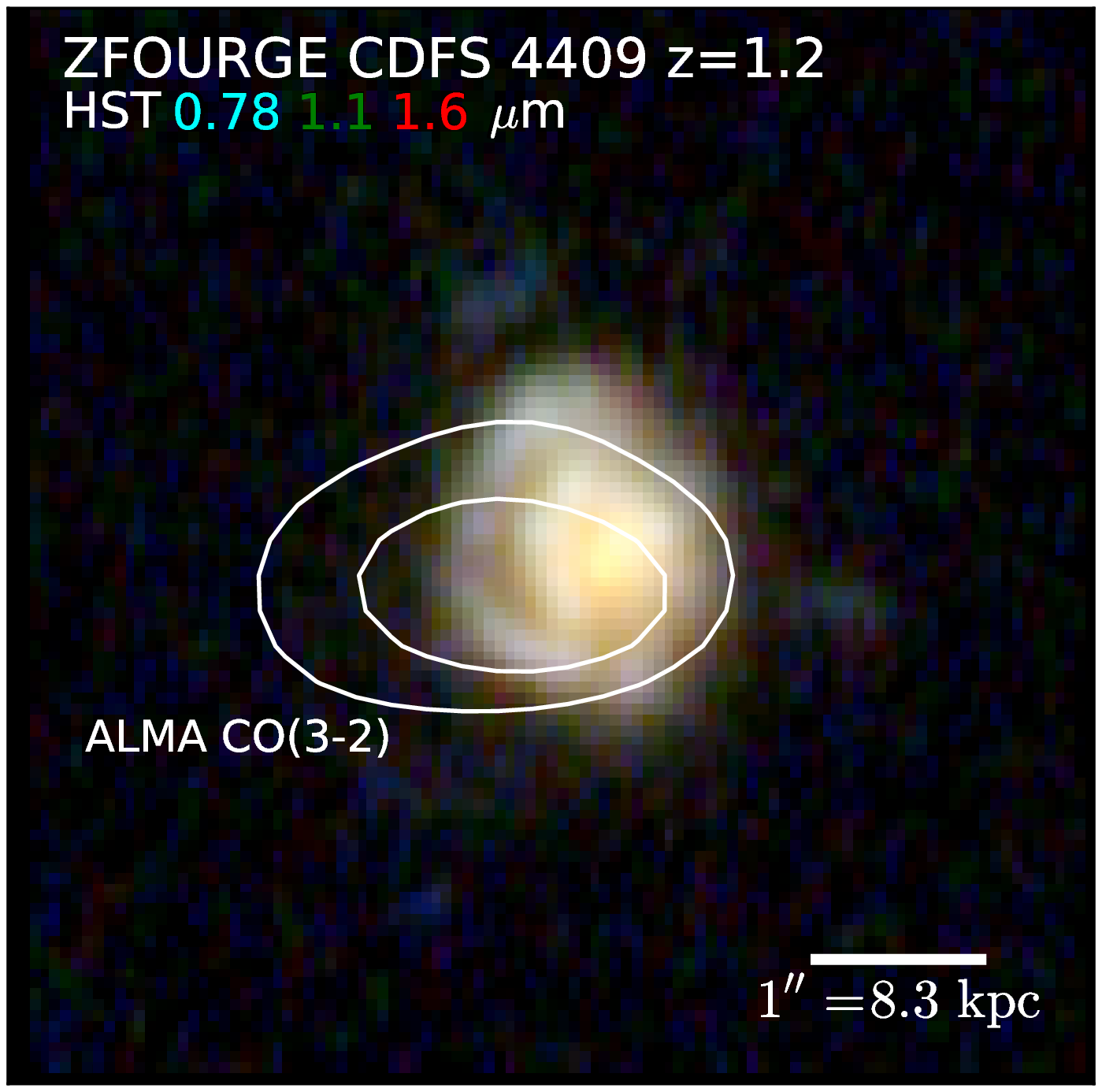}
        \includegraphics[width=0.25\textwidth]{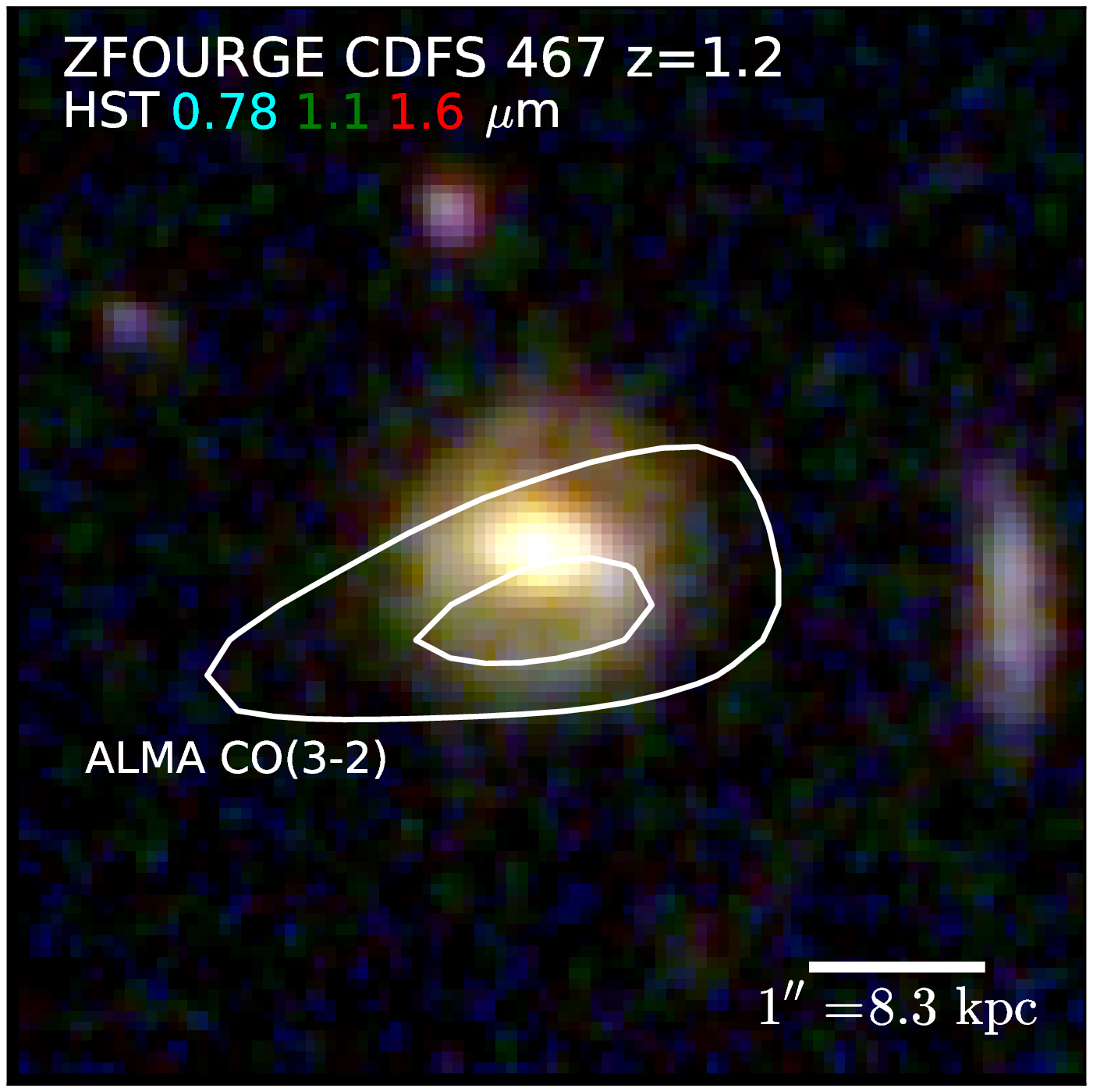}
      } 
      \ifpreprint \figureOneCaption \else \caption{} \fi
        \ifpreprint \vspace{-4mm} \fi
        \ifpreprint\end{figure*} \else \end{figure} \fi
}

\def\figureTwoCaption{
\caption{\textbf{Star Formation Efficiency as a function of CO
    luminosity, $\mathbf{L^\prime(CO)}$.}  The star-formation efficiency is defined as
  the ratio of the total IR luminosity (\lir) to \lco, where \lco\ is
  converted to the emission of the $J$=1--0 transition.  The 
  {$z=1.2-1.3$ galaxies in our sample} are shown as large red
  circles.  Error bars denote $1\sigma$ uncertainties.
  Other small symbols denote control samples of star-forming galaxies,
  including local spiral galaxies (open triangles), local
  ultraluminous IR galaxies (ULIRGS;
  crosses),\cite{gao04,combes11,combes13}  high
  redshift ($z > 1$) star-forming galaxies (cyan-filled squares), and high
  redshift submillimetre galaxies (yellow-filled
  diamonds).\cite{cari13,tacc13} The shaded regions show the
  interquartile ranges of the star-formation efficiency for local normal
  spirals and ULIRGs.\label{fig:lcolir} }
  }

\def\figureTwo{
  \setcounter{figure}{1}
  \ifpreprint \begin{figure}[ht]  
   \centerline{
     \includegraphics[width=0.5\textwidth, trim=4pt 10pt 5pt 5pt, clip=true]{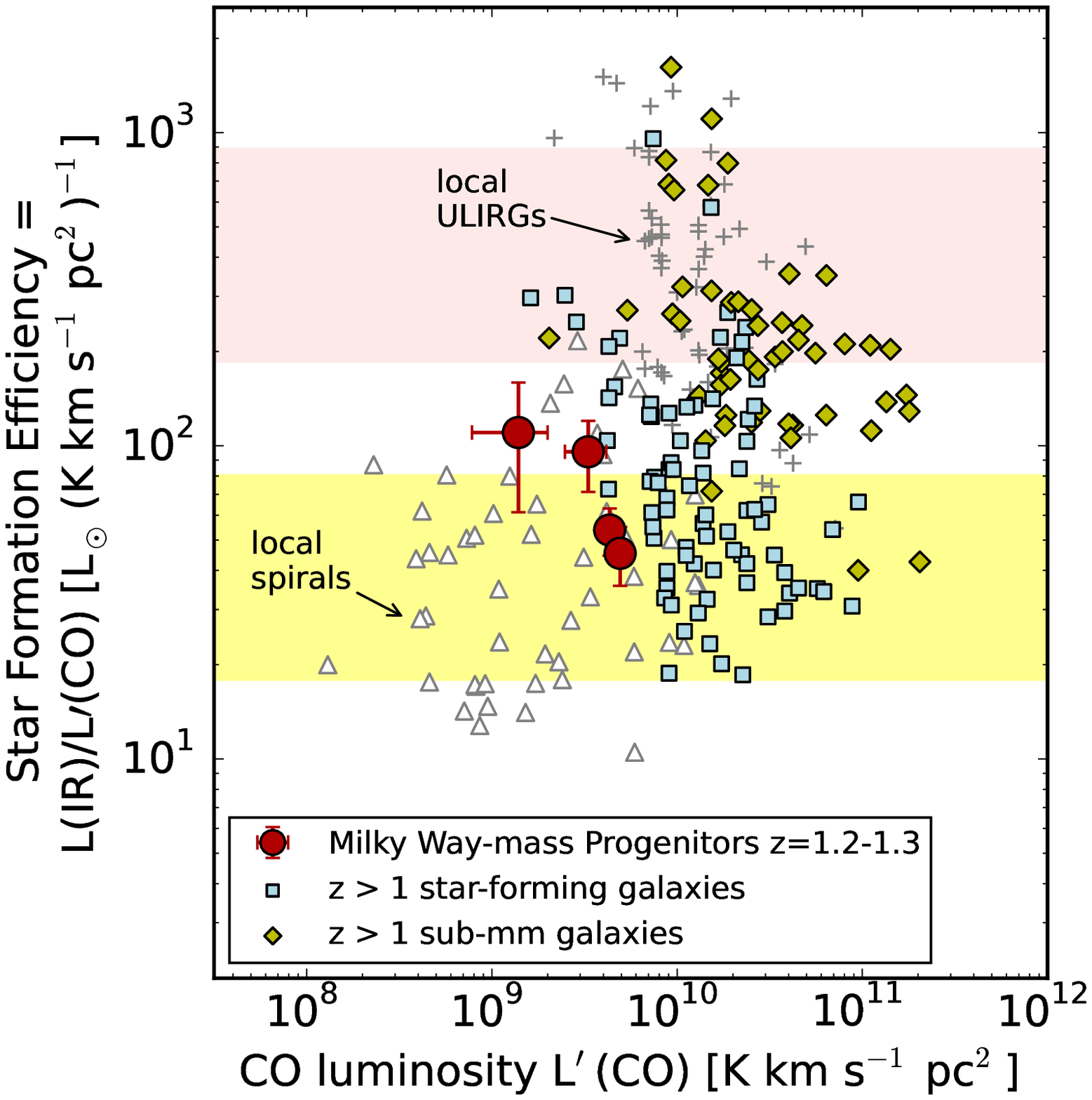}
   }
    \else \begin{figure} 
   \centerline{
     \includegraphics[width=0.85\textwidth, trim=4pt 10pt 5pt 5pt, clip=true]{lco_ratio.ps}
   }
   \fi
   \ifpreprint \figureTwoCaption \else \caption{} \fi
   \ifpreprint\vspace{-2mm} \fi 
 \end{figure}
}

\def\figureThree{
\setcounter{figure}{2}
  \ifpreprint  \begin{figure}[t] 
  \centerline{
    \includegraphics[width=0.5\textwidth, trim=4pt 10pt 5pt 5pt,
    clip=true]{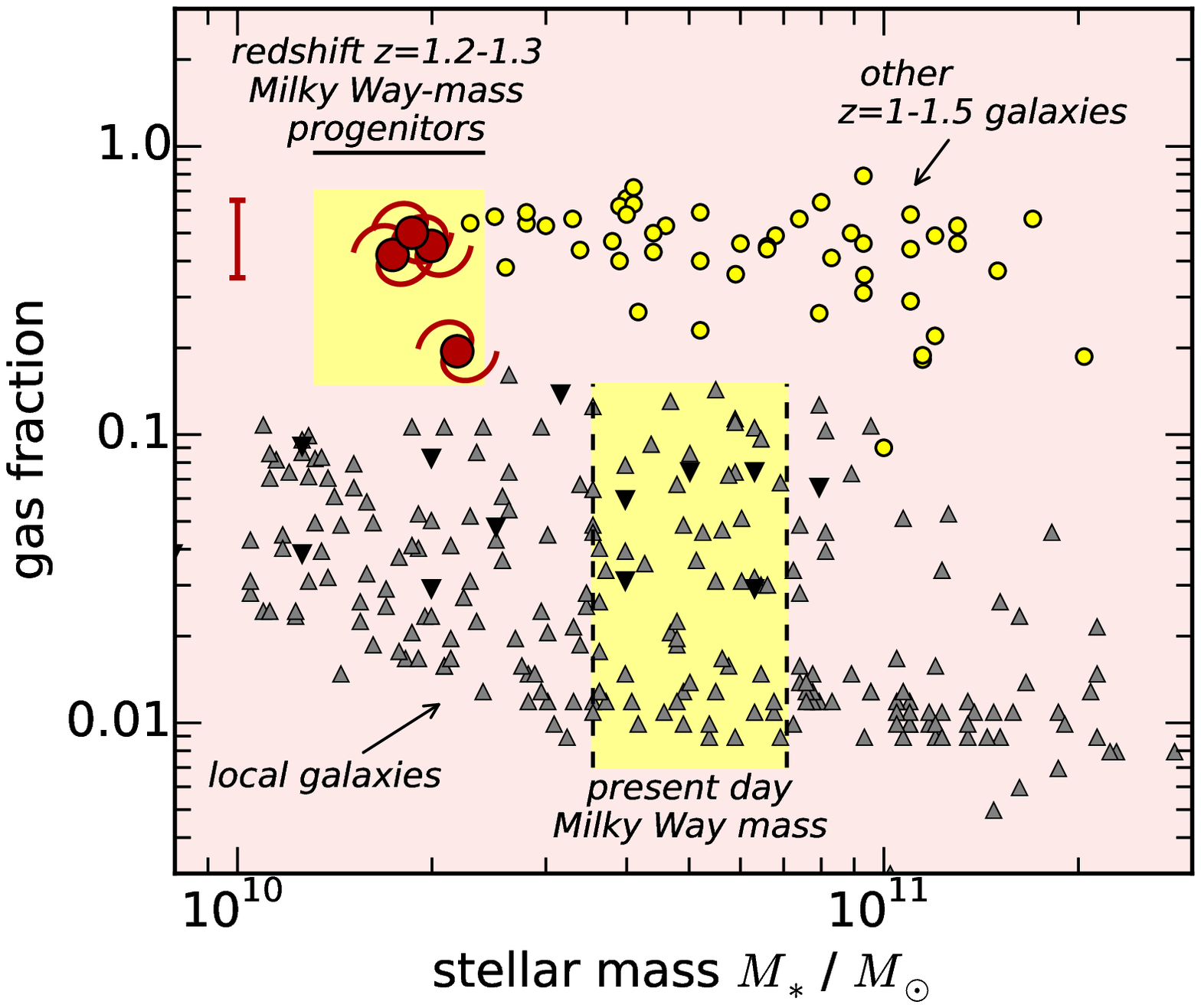}
  }
\else \begin{figure} 
  \centerline{
    \includegraphics[width=0.85\textwidth, trim=4pt 10pt 5pt 5pt,
    clip=true]{Fgas_onepanel.ps}
  }
\fi
  \ifpreprint \vspace{-2mm} \figureThreeCaption \else \caption{} \fi
  \ifpreprint\vspace{-2mm} \fi 
\end{figure}
}

\def\figureThreeCaption{
  \caption{\textbf{The relation between the molecular gas fraction and
      total stellar mass in galaxies at $\mathbf{z=1 - 1.5}$ compared to
      local galaxies.}   Here the gas fraction is defined as the ratio
    $M_\mathrm{gas} / (M_\mathrm{gas} + M_\ast)$.     The progenitors of
    Milky Way-mass galaxies at $z=1.2-1.3$ are denoted by large, red
    spirals. {The red bar shows the typical statistical
      uncertainty, $\approx$30\%.}.  Smaller, yellow circles show  other
    galaxies at $z=1 - 1.5$.\cite{daddi10a,magn12,tacc13} The smaller
    triangles show measurements for local ($z\sim 0$) galaxies with data
    from the literature, including COLD GASS\cite{sain11} (gray, upward
    triangles) and the HERA CO line excitation survey\cite{leroy09}
    (black, downward triangles).   The region separated by vertical dashed
    lines shows the stellar mass range of Milky Way--like galaxies at
    present.\label{fig:fgas}} 
}

\def\tableOne{
\ifpreprint
\else
\begin{landscape}
\fi
  \begin{table*}
    \caption{
      Properties of Progenitors of Milky-Way-Mass Galaxies at $z=1.2-1.3$}\label{table:main}
    \begin{center}
      \begin{tabular}{lccccccccc}
        \hline
        ZFOURGE & $z_\mathrm{opt}$$^a$ & $z_\mathrm{CO}$$^b$ & R.A.$^c$ &
        Decl.$^d$ & $I_{\mathrm{CO}(3-2)}$$^e$ & $L^\prime_{\mathrm{CO}}$$^f$
        & $M_\mathrm{gas}$$^\dag$ & $\lir^g$ & $M_\ast$$^\ddag$ \\
        ID   &   &  &  (deg.)  & (deg.)  & (Jy km s$^{-1}$) & (10$^9$ K km s$^{-1}$
        pc$^2$) & ($10^{10}$ \msol) &  ($10^{11}$ \lsol) & (10$^{10}$~\msol) \\
        \hline
        CDFS 467\phn & 1.220 & 1.221 & 53.05850 & $-$27.85678 &
        0.11(0.05) &  1.4(0.6) &  0.55(0.19)   &  
        1.5(0.1) & 2.2$^{+0.4}_{-0.8}$ \\
        CDFS 4409       & 1.220 & 1.220 & 53.18124 & $-$27.76566 & 
        0.25(0.06) &  3.3(0.8) & 1.4\phn(0.4\phn) & 
        3.1(0.1) & 1.7$^{+0.3}_{-0.3}$ \\
        CDFS 6497       & 1.215 & 1.215 & 53.04564 & $-$27.72493 &
        0.33(0.04) & 4.3(0.6) & 2.3\phn(0.3\phn) & 
        2.3(0.3) & 2.0$^{+0.3}_{-0.5}$ \\
        CDFS 8193       & 1.326 & 1.326 & 53.07405 & $-$27.69459 &
        0.31(0.05) & 4.9(0.8) & 2.0\phn(0.4\phn) & 
        2.2 (0.3) & 1.9$^{+0.1}_{-0.2}$ \\
        \hline
      \end{tabular}
    \end{center}
    Numbers in parentheses are $1\sigma$ uncertainties. $^a$Redshift from
    optical spectroscopy.\cite{vanz08} $^b$Redshift measured from ALMA
    CO(3--2) data (see Methods).  $^c$Right Ascension and
    $^d$Declination (J2000).  $^e$CO(3--2) flux density.  {$^f$CO
      luminosity, converted to the CO $J$=1 to 0 transition assuming
      $r_{31}$=0.66.\cite{bola15}}  $^\dag$Total molecular gas mass in $H_2$
    (see Methods).  $^g$Total IR luminosity from
    8--1000~$\mu$m (see Methods).  $^\ddag$Stellar mass
    measurements (see Methods).   
  \end{table*}
\ifpreprint
\else
\end{landscape}
\fi
}

\title{Large Molecular Gas Reservoirs in Ancestors of Milky Way-Mass Galaxies 9 Billion Years Ago}

\author{C.~Papovich$^{1,2}$, I.~Labb\'e$^3$, K.~Glazebrook$^4$, R.~Quadri$^{1,2}$,
  G.~Bekiaris$^4$,
M.~Dickinson$^5$, S.~L.~Finkelstein$^6$, D.~Fisher$^4$, H.~Inami$^{5,7}$,
R.~C.~Livermore$^6$,  
L.~Spitler$^{8,9}$,  C.~Straatman$^3$, K.-V.~Tran$^{1,2}$}

\begin{document}

\maketitle

\let\thefootnote\relax\footnote{
\begin{affiliations}
\item George P. and Cynthia Woods Mitchell Institute for Fundamental
  Physics and Astronomy, Texas A\&M University, College Station, Texas
  78743, USA
\item Department of Physics and Astronomy, Texas A\&M University, 4242
  TAMU, College  Station, TX 78743, USA
\item Leiden Observatory, Leiden University, P.O. Box 9513, 2300 RA
  Leiden, The Netherlands
\item Centre for Astrophysics \& Supercomputing, Swinburne University,
  Hawthorn, VIC 3122, Australia
\item National Optical Astronomy Observatory, Tucson, AZ 85719, USA
\item Department of Astronomy, The University of Texas at Austin, 2515
  Speedway, Stop C1400,  Austin, TX 78712, USA
\item Centre de Recherche Astrophysique de Lyon, Universit\'e de Lyon,
  Universit\'e Lyon 1, CNRS, Observatoire de Lyon, 9 avenue Charles
  Andr\'e, F-69561 Saint-Genis Laval Cedex, France
\item Department of Physics \& Astronomy, Macquarie University,
  Sydney, NSW 2109, Australia
\item Australian Astronomical Observatory, PO Box 915, North Ryde, NSW 1670, Australia
\end{affiliations}
}

\vspace{-3.5mm}
\begin{abstract} 
{The gas accretion and star-formation histories of galaxies like the
Milky Way remain an outstanding problem in
astrophysics.\cite{guedes11,martig12}   Observations show that 8
billion years ago, the progenitors to Milky Way-mass galaxies were
forming stars 30 times faster than today and predicted to be rich in
molecular gas,\cite{papo15} in contrast with low present-day gas
fractions ($<$10\%).\cite{sage93,leroy09,sain11}   Here we show
detections of molecular gas from the CO($J$=3--2) emission (rest-frame
345.8~GHz) in galaxies at redshifts $\mathbf{z}$=1.2--1.3, selected to
{have the stellar mass and star-formation rate of the
progenitors} of today's Milky Way-mass galaxies.  The CO emission
reveals large molecular gas masses, comparable to
or exceeding the galaxy stellar masses, and implying most of the
baryons are in cold gas, not stars.   The galaxies'
total luminosities from star formation and CO luminosities yield
long gas-consumption timescales.  {Compared to local spiral galaxies,}
the star-formation efficiency, {estimated from the ratio of total IR
luminosity to CO emission,}  has remained nearly constant since
redshift $\mathbf{z}$=1.2, despite the order of magnitude decrease in
gas fraction, consistent with results for other galaxies at this
epoch.\cite{daddi10a,magdis12,magn12,tacc13}  Therefore {the physical
processes that determine the rate at which gas cools to form stars in
distant galaxies appear to be similar to that in local galaxies}.}
\end{abstract}

Studies of the distribution of stellar ages and elemental abundances
in the Milky Way and M31 conclude most of their stars  formed in the
distant past, more than 7 billion years ago.\cite{snaith14,bern15}
This agrees with recent work that shows star-formation in present-day
galaxies with the mass of the Milky Way peaked more than 8 billion
years ago, at $z$$>$1,\cite{papo15} with star-formation rates (SFRs)
exceeding 30 \msol\ yr$^{-1}$,  compared to a present day {SFR of
1.7$\pm$0.2 \msol\ yr$^{-1}$ for the Milky Way.\cite{licq15}}   

Theoretical models explain periods of high star formation as a result
of rapid baryonic gas accretion from the intergalactic medium (IGM),
which leads to high cold gas concentrations in galaxies at earlier
times.\cite{delucia14}  {{These models predict that the
  gas settles} into rotationally
supported, highly turbulent disks, which fragment to form
stars.\cite{dekel09a} Observations of star-forming
galaxies at $z$$>$1 (stellar masses, $M_\ast$$>$$2$$\times$$10^{10}$~\msol) show
evidence for gas-rich, rotating disks,\cite{genzel06,tacc10,genzel15,wisn15}
supporting these theories.}  However, the situation is far from settled
for lower mass ($M_\ast \sim 10^{10}$~\msol), more common galaxies
such as the progenitors to the Milky Way.   Some models predict that
these galaxies should experience early, rapid star formation, leaving
low gas fractions ($<$10\%) at redshifts $z$$\sim$1.\cite{ceve10}
Others predict that the gas flows from the IGM can perturb and disrupt
the formation of disk instabilities, thereby suppressing star
formation in galaxies and extending star-formation
histories.\cite{genel12,gabor14}   The first step to understand star
formation in galaxies like the Milky Way is to measure the amount of
the cold gas in their progenitors at $z$$>$1.  As the gas is the fuel
for star formation, the ratio of the SFR to gas mass can test the physical
processes in the models.\cite{agertz15}

With the greatly improved sensitivity offered by the Atacama Large
Millimeter Array (ALMA), we are able now to explore the evolution of
cold molecular gas in low mass galaxies at redshifts $z>1$.  {With
ALMA, we observed the $J$=3 to 2 transition of CO in four galaxies
{with the stellar mass and SFR expected of the main
progenitors to present-day Milky Way-mass galaxies at redshifts
$z=1.2-1.3$ selected from deep imaging by the FourStar Galaxy
Evolution (ZFOURGE) survey\cite{stra16}  (see discussion in the
Methods).}}  Figure~\ref{fig:co_hst} shows the
integrated emission from the CO $J=3$ to 2 transition in these
galaxies, {where the detections range in significance from
4.8--13.7$\sigma$ (r.m.s.).}  The CO(3--2) emission coincides with the
spatial positions of the galaxies in  Hubble Space Telescope (HST)
imaging (Figure~\ref{fig:co_hst}); the small offsets are consistent
with astrometric calibrations and ALMA beam smearing.
Table~\ref{table:main} gives the measured properties of these
galaxies.   {{The ALMA detections of CO emission probe the
molecular mass in galaxies with the stellar mass and SFRs that the
main progenitor of the Milky Way was expected to have $\sim$8.5 billion
years ago.   This provides an important extension of previous work, as
the galaxies in our sample have lower stellar masses and SFRs than have been
generally possible to study at these redshifts.\cite{tacc13}}}

\ifpreprint \figureOne \fi
\ifpreprint \tableOne \fi

CO is the most luminous tracer of molecular hydrogen ($H_2$), the fuel
for star formation.  The CO specific intensity from the $J$ to $J$--1
transition, $I_{\mathrm{CO}{(J-[J-1])}}$, { is a function of both the
gas density and temperature.  {In high redshift galaxies, studies show
that the average excitation of CO(3--2) is similar to that of
  star-forming regions in the Milky Way},\cite{daddi15} and we assume an integrated
Rayleigh--Jeans brightness temperature line ratio, $r_{31}$ =
$I_{\mathrm{CO}(3-2)} /I_{\mathrm{CO}(1-0)} \times (1/3)^2$ =
0.66.\cite{bola15}} The total CO luminosity in the $J$=1 to 0
transition is then $L^\prime_\mathrm{CO}$ = $3.25 \times 10^7\
r_{31}^{-1} \ I_{\mathrm{CO}(3-2)}\ \nu_\mathrm{obs}^{-2}\ D_L^2\
(1+z)^{-3}$ where $\nu_\mathrm{obs}$ is the frequency (in GHz) of the
CO emission in the observed frame and $D_L$ is the luminosity distance
in Mpc.   Table~\ref{table:main} displays the \lco\ values.  {Using
lower values of $r_{31}\sim 0.4-0.5$, as indicated by some other
studies of star-forming galaxies at $z$$\sim$1--2,
\cite{arav14,daddi15} would increase the \lco\ values slightly but not
change our conclusions.}

\ifpreprint\figureTwo\fi

The combination of the CO luminosity and the luminosity from newly
formed stars provides a crucial constraint on the star-formation
efficiency (SFE).   We use the thermal IR luminosity (\lir, measured
over 8--1000~$\mu$m in the rest frame), which originates from dust in
dense molecular clouds heated by young stars, and is directly
proportional to the total SFR.    We measured \lir\ for galaxies in
our study using model fits to fluxes measured from \textit{Spitzer
Space Telescope} and \textit{Herschel Space Observatory} imaging
covering 24--250~$\mu$m (see Methods).
Table~\ref{table:main} displays these values.  {They span
\lir=$(1.5-2.7)$$\times$$10^{11}$~\lsol\ (corresponding to SFRs of
15--30~\msol\ yr$^{-1}$).  Uncertainties are $\simeq$0.2 dex (60\%)
and are dominated by systematics from differences in the IR model (see
Methods).}

Figure~\ref{fig:lcolir} shows the SFE, defined as \lir/\lco, as a
function of \lco\ for the $z$=1.2--1.3 {galaxies in our sample}
compared to control samples.  {With ALMA we now probe
efficiently the  CO luminosities of $z > 1$ star-forming
galaxies a factor two lower than was possible previously.   The
{galaxies in our sample} have SFEs typical of the upper
range of both local spiral galaxies and more massive, high-redshift
star-forming galaxies.}  In such galaxies, star-formation occurs in
rotationally supported disks.  {In at least two of our galaxies, the
CO(3--2) spectra show strongly double--peaked line profiles (see
Methods).  This and the apparent presence of spatial
velocity shear (see Methods) observed in our analysis
of the CO data  suggests that the same may be true for all the
$z$=1.2-−1.3 {galaxies in our sample}.}   Therefore, even though both
the SFRs and gas fractions are substantially higher in these distant
galaxies, their star formation likely occurs in rotating disks,
{where the physical processes governing the evolution of the
  gas appears to be similar to that of} spiral galaxies in the local Universe. In contrast,
the SFEs of more luminous, rarer objects (ULIRGs, QSOs, and
submillimetre galaxies [SMGs]) are significantly enhanced in the
local and distant Universe.  A prevailing theory is that ULIRGs,
QSOs, and SMGs are a result of increased gas densities from major
gas-rich mergers.\cite{casey14} These conditions seem inconsistent
with the {galaxies in our sample, suggesting that major
  mergers are not common amongst the main progenitors of Milky
  Way-mass galaxies at $z$$\simeq$1.2--1.3.}

The inverse of the SFE is proportional to the gas consumption
timescale, which corresponds to a range of 200 to 700~Myr for the
{galaxies in our sample}.   In contrast, the consumption
timescales for ULIRGs, QSOs, and SMGs are less than 10
Myr.\cite{cari13} {Star-formation in the average, main progenitor of Milky Way
galaxies at $z$=1.2--1.3 appears to be long lasting}{, and comparable to
findings for other star-forming disk galaxies at high
redshifts.\cite{daddi10a,magdis12,magn12,tacc13,genzel15}}

The CO luminosities imply very high molecular gas fractions for {the
galaxies in our sample at $z$=1.2--1.3}, where we adopt the
ratio of CO luminosity to mass in $H_2$ gas ($M_\mathrm{gas}$) for Galactic star-forming
regions because the SFEs are similar (see Methods).
Table~\ref{table:main} lists these values.   Figure~\ref{fig:fgas}
shows the molecular gas fractions ($f_\mathrm{gas} = M_\mathrm{gas}$ /
$(M_\mathrm{gas} + M_\ast)$) {derived from CO observations} as a function of the stellar mass,
$M_\ast$.    While present day Milky Way--sized galaxies have low gas
fractions, $f_\mathrm{gas} < 10$\%, {the results from our
  sample imply that the main progenitors to these galaxies at $z=1.2-1.3$
have much higher values:  in three of the galaxies in our sample}, {the
molecular gas mass is greater than or equal to the stellar mass ($f_\mathrm{gas} \gtrsim
50$\%).   This is consistent with indirect gas fractions of galaxies
at these redshifts inferred from the thermal dust emission.\cite{genzel15,scov16}}
This argues against models with early, rapid gas consumption
\cite{ceve10} and favours longer lasting, feedback--regulated
star-formation.\cite{genel12,gabor14,agertz15} 

\ifpreprint\figureThree\fi

The high molecular gas fractions and SFRs of the $z$=1.2--1.3
{galaxies in our sample} imply they will double their
stellar mass within the gas-consumption timescale.     Therefore, at
$z$$\sim$1.2 these galaxies have most, but not all, of the fuel needed
to produce the $M_\ast \simeq 5\times 10^{10}$~\msol\ in stars  in
Milky Way-mass galaxies at present (see Figure~3).  The
average baryon accretion rate from the IGM must exceed 6~\msol\
yr$^{-1}$ at earlier times ($z$$>$1.2) to account for the galaxies'
total stellar and molecular masses.  In constrast, the galaxies need
only acquire $\sim$30--50\% more baryonic mass from $z$$\sim$1 to the
present (even accounting for losses from stellar evolution), which
corresponds to an average gas accretion rate of only $\sim$1--2~\msol\
yr$^{-1}$.  This reflects a dwindling supply of fresh baryonic gas.
Therefore, Milky Way-mass galaxies appear to accrete most of their gas
at $z$$>$1.2, during the first few billion years of history.


\begin{addendum}
\item   [Acknowledgements]   The authors thank their colleagues on the
CANDELS and ZFOURGE surveys, for providing high quality data products.
The authors thank the ALMA staff for facilitating the observations,
and aiding in the calibration and reduction process.  {The authors
  also wish to acknowledge the anonymous reviewers for their careful
  reading, valued comments, and constructive criticism.} The authors
acknowledge generous support from the Mitchell Institute for
Fundamental Physics and Astronomy at Texas A\&M University.    This
paper makes use of the following ALMA data:
ADS/JAO.ALMA\#2011.0.01234.S . ALMA is a partnership of ESO
(representing its member states), NSF (USA) and NINS (Japan), together
with NRC (Canada), NSC and ASIAA (Taiwan), and KASI (Republic of
Korea), in cooperation with the Republic of Chile. The Joint ALMA
Observatory is operated by ESO, AUI/NRAO and NAOJ. The National Radio
Astronomy Observatory is a facility of the National Science Foundation
operated under cooperative agreement by Associated Universities, Inc.

\item[Author  Contributions]   C.~Papovich led the ALMA observing
programme, handled the data reduction, and led the writing of the text
of the manuscript.  Coauthors I. Labb\'e, K.~Glazebrook, R.~Quadri,
L.~Spitzer, C.~Straatman, and K.-V.~Tran contributed extensively to
the ZFOURGE dataset, used in much of the analysis.  Coauthors
S.~Finkelstein, D.~Fisher, and R.~Livermore contributed to the design
of the ALMA observing programme, and assisted in the ALMA data reduction
and interpretation.  Coauthors G.~Bekiaris and K.~Glazebrook assisted
in the interpretation of the ALMA data.  Coauthors
M.~Dickinson and H.~Inami carried out the data analysis of the
\textit{Spitzer} and \textit{Herschel} imaging.  All co-authors
contributed to the writing of the manuscript, and to the ALMA
observing programme.  

\item[Author  Information]  Reprints  and permissions  information  is
  available  at   www.nature.com/reprints.  The  authors   declare  no
  competing financial interests. Readers are welcome to comment on the
  online  version  of  the  paper.  Correspondence  and  requests  for
  materials should be addressed to C.P. (papovich@tamu.edu).
\end{addendum}

\ifpreprint \else 
\pagebreak
\begin{figure} \figureOneCaption \end{figure}
\vspace{18pt}

\begin{figure} \figureTwoCaption \end{figure}
\vspace{18pt}

\begin{figure} \figureThreeCaption \end{figure}
\pagebreak
\fi

\begin{addendum}
\item[Data Availability Statement]  The data that support the plots
within this paper and other findings of this study are available from
the corresponding author upon reasonable request. Data from the
ZFOURGE survey can be obtained from http://zfourge.tamu.edu/

\end{addendum}

\clearpage

\ifpreprint
\else
\setcounter{page}{1}
\fi
\setcounter{figure}{0}
\setcounter{table}{0}
\renewcommand{\thefigure}{\arabic{figure}}
\renewcommand{\thetable}{\arabic{table}}

\begin{center}
{\bf \Large \uppercase{Methods} }
\end{center}

\Supplementarytrue

\section{ZFOURGE Dataset}

We selected the {galaxies in our sample from the}
Fourstar Galaxy Evolution (ZFOURGE) survey\cite{stra16}.    The main
ZFOURGE survey obtained very deep near-infrared imaging in five
medium--band filters (\jone, \jtwo, \jthree, \hs, \hl) from the
FourStar instrument\cite{pers13} on the Magellan Baade 6.5~m telescope
in the three southern fields covered by CANDELS \textit{HST}
imaging.\cite{grogin11,koek11}    The ZFOURGE catalogues combine  the
FourStar images with ancillary ground-based imaging (spanning 0.3 --
2.5~$\mu$m), the CANDELS \textit{HST}/ACS and WFC3 imaging, and
Spitzer/IRAC imaging (spanning 3.6 -- 8.0~$\mu$m).   For this study,
we selected targets from the earlier, version 2.1 ZFOURGE catalogues.  These include
photometric redshifts and stellar masses estimated from the
multiwavelength catalogues as described elsewhere.\cite{tomc14,papo15}
Of interest here, the catalogues are complete for objects with limiting
stellar masses $\log M/\msol \geq 9.0-9.2$ in the redshift range $1.0
< z < 1.5$, well below the stellar masses of the typical progenitor of
a Milky-Way-mass galaxy.\cite{papo15}  

\section{Selection of Milky-Way-Mass Galaxy Progenitors}

{We selected galaxies as targets for ALMA observations of the
CO($J$=3 to 2) transition that  have the typical stellar mass and SFR
of progenitors to Milky Way-mass galaxies at $z=1.2-1.3$.}    We
identified progenitors of galaxies with the present-day stellar mass
of a Milky Way-mass  galaxy ($M_\ast = 5\times 10^{10}$~\msol\ at
$z=0$)\cite{mutch11,licq15} using abundance-matching
techniques.\cite{moster13}    The progenitors to such galaxies had a
median stellar mass $\log M_\ast / \msol = 10.21$ at
$z=1.1-1.4$.\cite{papo15}  {These abundance matching methods
give stellar mass $\approx$0.2 dex lower than those selected at
constant co-moving number density at these
redshifts.\cite{vandokkum13}  More recent work shows that progenitors
of Milky Way-mass galaxies span a range of stellar mass at $z=1.2$,
with a 30th to 70th-tile range of $\log M=10.05$ to 10.34, and a
median value consistent with the median above.\cite{well16}    While
observations of CO in $z > 1$ galaxies have probed stellar masses down
to $\log M_\ast/\msol > 10.4$,\cite{tacc13} these correspond to the
more massive progenitors of present-day Milky Way-mass galaxies.  Our
sample extends studies of the CO emission to the median stellar mass
of progenitors of present-day Milky Way galaxies.} 

{We also selected galaxies with the typical SFRs of the Milky Way-mass
progenitors for observations with ALMA.}  In our previous work we used
deep \textit{Spitzer} and \textit{Herschel} imaging to measure an
average total IR luminosity, $\lir = (2.0 \pm 0.1) \times
10^{11}$~\lsol, for all Milky Way-mass progenitor galaxies in this
redshift and stellar mass range in ZFOURGE.\cite{papo15}   This
corresponds to a SFR $= 21 \pm 2$~\msol\ yr$^{-1}$.   

In summary, we used the following criteria to select targets for ALMA:
\begin{compactenum}
\item Photometric redshift,  $1.1 < z < 1.4$;
\item Stellar mass,  $-0.15  <  \log M_\ast/\msol - 10.2 < +0.15$;
\item SFR,  $-0.15 < \log \mathrm{SFR}/\msol\ \mathrm{yr}^{-1} - 1.3 < +0.15$;
\item measured spectroscopic redshift.
\end{compactenum}
The restrictions on photometric redshift, stellar mass, and SFR select
galaxies with stellar mass and SFR within 0.15~dex (i.e., within 40\%)  of
the expected median values of the progenitors to Milky Way-mass galaxies.

\ifpreprint
\begin{figure}
\centerline{
\includegraphics[width=0.45\textwidth, trim=5pt 0pt 0pt 5pt, clip=true]{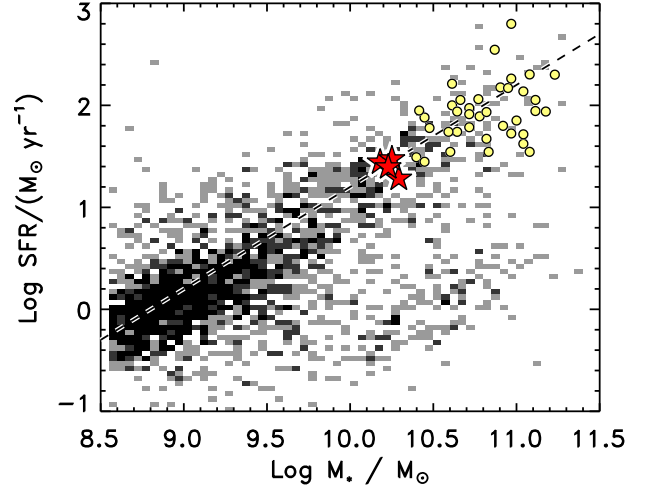}}
\vspace{-4mm}
\caption{\textbf{SFR--stellar mass sequence for ZFOURGE galaxies at
$1.1 < z < 1.4$.}  The shading increases with the number density of
galaxies in each bin.   The ``main sequence''  of star-formation is
indicated by the dashed line, and has a slope of SFR $\propto$
$M_\ast$.\cite{tomc16}   The large red stars indicate the four
{sources selected as typical of main progenitors of Milky Way-mass
galaxies at $z=1.2-1.3$ observed with ALMA here.}  {The yellow
circles indicate objects with $1.0 < z < 1.5$ with CO detections from PHIBSS.\cite{tacc13}}
}\label{fig:sfrms}  
\vspace{-1mm}
\end{figure}
\fi 

\ifpreprint
\begin{figure}
\centerline{
\includegraphics[width=0.5\textwidth, trim=5pt 0pt 0pt 5pt, clip=true]{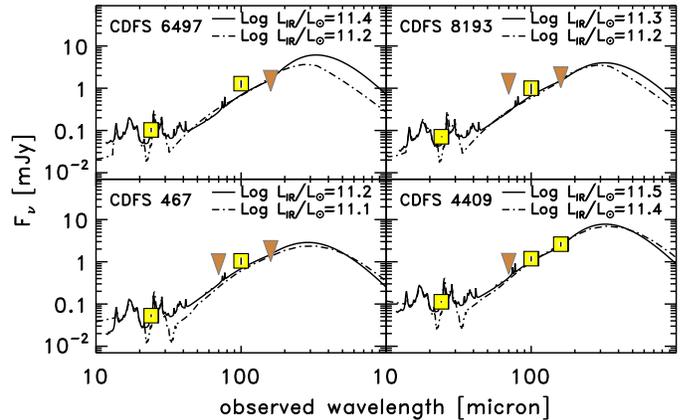}
}
\vspace{-4mm}
\caption{\textbf{IR Spectral Energy Distribution Model Fits to the
far-IR data. }    The yellow squares show the \textit{Spitzer}
24~$\mu$m, and \textit{Herschel} 70, 100, and 160~$\mu$m flux
densities measured for each object.   Error bars are $1\sigma$
uncertainties.  {Downward triangles indicate $3\sigma$ upper limits for
sources detected with S/N $<$ 3 at that wavelength.}  The curves show
model fits for the Rieke et al.\cite{rieke09} (solid lines) and Chary
\& Elbaz\cite{chary01} (dot--dashed lines), which bracket the range of
values for total IR luminosity, \lir.  }\label{fig:lir}
\vspace{-5mm}
\end{figure}
\fi

\ifpreprint
{
\begin{table}
\begin{center}{\bf Supplementary Table 1: Summary of Far-IR Flux Densities}\end{center}
\begin{center}
\begin{tabular}{lcccc}
\hline
\hline
ZFOURGE ID: &                                               467
&  4409  & 6497  & 8193 \\ \hline
$F_\nu(24\mu\mathrm{m})$: &  53 (4) &  113 (5) &
103 (12) & 71 (12) \\ 
$F_\nu(70\mu\mathrm{m})$: & $<$1.2     &  $<$0.9    &
\ldots  & $<$1.4  \\ 
$F_\nu(100\mu\mathrm{m})$: &  1.1 (0.2)  & 1.2 (0.2)    & 1.3
(0.3) & 1.0 (0.3) \\ 
 $F_\nu(160\mu\mathrm{m})$: &  $<$2.0    & 2.6 (0.3)    & $<$1.7   & $<$2.1 \\
\hline
\end{tabular}
\end{center} Numbers in parentheses are $1\sigma$ uncertainties.
24~$\mu$m flux densities are in units of $\mu$Jy.  All other flux
densities are in units of mJy, where 1 $\mu$Jy = $10^{-3}$ mJy, and 1
mJy = $10^{-26}$~erg s$^{-1}$ cm$^{-2}$ Hz$^{-1}$.  Upper limits are
$3\sigma$.  \vspace{0mm}
\end{table}
}
\fi

\ifpreprint
\begin{table*}
\begin{center}{\bf Supplementary Table 2: Summary of ALMA observations}\end{center}
\begin{center}
\begin{tabular}{lccccccc}
\hline
\hline
ZFOURGE&  Obs.\ Dates & PMV$^\mathrm{b}$ &  $T_\mathrm{int}$$^\mathrm{c}$  & Frequency Range & Combined
Beam$^\mathrm{d}$ &  FWHM$^\mathrm{e}$ &  $\sigma_\mathrm{r.m.s.}^\mathrm{f}$ \\
ID$^\mathrm{a}$     &     & (mm)  & (min) & (GHz)   & & (km s$^{-1}$) &  (mJy beam$^{-1}$)  \\ \hline
CDFS 467 (8769) &  2015 Apr 7 & 3.2 & 37.3 &  140.9--156.6 &  $3\farcs5
\times 1\farcs3$, P.A. = $-$74.1$^\circ$\phantom{$^\mathrm{,g}$} & 240 (120)\phantom{$^\mathrm{g}$} & 0.42\phantom{$^\mathrm{g}$} \\ \hline
CDFS 4409 (19996) & 2015 Apr 7 & 3.6 & 37.3 &  140.9--156.6 &
$2\farcs4\times 1\farcs4$, P.A. = $-85.8^\circ$\phantom{$^\mathrm{,g}$} & 440 (150)\phantom{$^\mathrm{g}$}   & 0.42\phantom{$^\mathrm{g}$} \\ \hline
CDFS 6497 (24956) & 2015 May 2 & 1.0 & 37.8 &  141.3--156.9 & $2\farcs7
\times 1\farcs3$, P.A. = $-77.9^\circ$$^\mathrm{,g}$ &  310 \phantom{1}(50)$^\mathrm{g}$ & 0.33$^\mathrm{g}$  \\
         & 2015 May 2 & 1.5 & 37.8 &  141.3--156.9 & &  &  \\ \hline 
CDFS 8193 (28279) & 2015 Apr 6 &  4.6 & 41.8 &  134.5--149.5 & $2\farcs 8 \times
1\farcs4$, P.A. = $-$81.7$^\circ$\phantom{$^\mathrm{,g}$} &  \phantom{1}54 \phantom{1}(16)\phantom{$^\mathrm{g}$}  & 0.48\phantom{$^\mathrm{g}$} \\
\hline
\end{tabular}
\end{center} $^\mathrm{a}$Source ID in the previous ZFOURGE v2.1 proprietary catalogues used
to select targets for ALMA.  {The numbers in parentheses are the ZFOURGE
source ID in the public v3.4 catalogues.\cite{stra16}}  $^\mathrm{b}$Estimated precipitable water vapour.
$^\mathrm{c}$On-source integration
time. $^\mathrm{d}$FWHM and orientation of cleaned beam (numbers in
parentheses are 1$\sigma$ uncertainties).  These are
displayed graphically in figure~\ref{fig:co_hst}. $^\mathrm{e}$The
FWHM of the CO(3-2) line in the measured spectrum.
$^\mathrm{f}$R.M.S. noise per channel measured in primary-beam
corrected data. $^\mathrm{g}$For this object, the beam size, FWHM, and noise
are measured from the combined, 2-epoch dataset.\vspace{0mm}
\end{table*}
\fi

The final selection criteria requires that the galaxies have a redshift
measured from spectroscopy.  This ensures that the redshift accuracy is
sufficient for the redshifted CO(3--2) emission line to fall within the
frequency range of an ALMA spectral window.  While the ZFOURGE photometric
redshifts are good ($\sigma_z/(1+z) < 1$\%)\cite{stra16}, these are
not sufficient for this purpose.      

Of the 24,690 galaxies in the full ZFOURGE catalogue, 39 satisfied the
first three selection criteria.   At the time of our proposal for ALMA
for cycle 2 observations (2013 December), 7 galaxies satisfied all our
selection criteria (including having a published spectroscopic
redshift in the literature).\cite{vanz08}  From these, we selected
four objects offering some contrast in SFR (spanning nearly 0.3~dex).
{For the analysis in this Letter, we rederived stellar masses and uncertainties
using FAST \cite{kriek09b} with an extended stellar population library
(including a broader  metallicity range of $0.2-1.0$~$Z_\odot$ and a
finer grid spacing of star-formation histories) compared to the one
used for the ZFOURGE catalogues.\cite{stra16}    These stellar
masses and 68\% likelihood range are listed in Table~\ref{table:main},
and are consistent with those in the v2.1 and v3.4 ZFOURGE catalogues. } 

All four galaxies selected for ALMA observations have properties
characteristic of the typical properties of progenitors to a Milky
Way-mass galaxy at $z=1.2-1.3$.     Supplementary Figure~\ref{fig:sfrms} shows the
SFR--stellar mass relation for galaxies from ZFOURGE with $1.1 < z <
1.4$.   Galaxies with the stellar mass and  and SFR of the Milky
Way-mass progenitors lie on the star-forming ``main
sequence'',\cite{tomc16} and this includes the four galaxies we
observed with ALMA.  Therefore, they correspond to a typical
star-forming galaxy at these redshifts.    {Previous studies of CO
emission in galaxies at these redshifts have been limited to higher
SFRs ($\gtrsim$30~\msol\ yr$^{-1}$) and/or stellar masses
($M_\ast$$>$$2.5$$\times$$10^{10}$~\msol)\cite{magn12,tacc13}.  As
illustrated in Supplementary Figure~\ref{fig:sfrms},  {the observations of
our sample} with ALMA provide an important extension compared to
previous studies.  Furthermore, previous studies required up to 25 hrs
of integration with IRAM PdBI to detect galaxies at these mass and SFR
limits.\cite{tacc13}  Our  ALMA observations required only $\simeq$40
min, demonstrating the efficacy of ALMA for this science.}

\section{Far--IR data and IR Luminosities} 

The ZFOURGE fields include imaging at far--IR wavelengths from
{\textit{Spitzer}/MIPS (24~$\mu$m), \textit{Herschel}/PACS
(70, 100, and 160~$\mu$m).}
Fluxes are measured in these data using source detections based on
prior locations of sources in \textit{HST}/WFC3 F160W (1.6~$\mu$m)
data using methods identical to those described
elsewhere.\cite{elbaz11,magn11,magn13}   {We measured flux
uncertainties  and evaluated source completeness through extensive
artificial object simulations, following the same procedures discussed elsewhere\cite{magn13,berta13}}   The {24--160~$\mu$m} flux
densities for the four objects studied here are listed in Table~S1.
(Note that one source, ZFOURGE  CDFS 4409, has no coverage by PACS
70~$\mu$m).  

{In all cases,  the IR flux densities and flux uncertainties that we
  measure for our sources are consistent with other 
published values,\cite{elbaz11,lutz11} available on the WWW
(http://irsa.ipac.caltech.edu/data/Herschel/PEP).}   {In many cases, our
measured flux densities at 70~$\mu$m and 160~$\mu$m have
S/N$<$3}.  While formally undetected, we include this information in
our analysis as it provides important constraints on the total IR
emission from these galaxies. 

 To measure total IR luminosities, $\lir$, we fit models of the IR
spectral energy distribution\cite{chary01,dale05,rieke09} to the flux
densities in Table~S1.   Because the data sample well the Wein side
of the thermal emission, the constraints on the IR luminosity are
quite robust.  Supplementary figure~\ref{fig:lir} shows the fits using the Chary \&
Elbaz and Rieke et al.\ models,\cite{chary01,rieke09} which bracket
the range of values.  The slight differences in IR spectral energy
distribution shape lead to systematically different IR luminosities,
where \lir\ from the Rieke et al.\ templates are higher by
$\Delta(\log \lir)$ = 0.1--0.2 dex.   { We have also calculated IR
luminosities ignoring data where objects are detected at $<$2$\sigma$,
but this produces changes in the IR luminosities by $<$15\% in most
cases.}  We therefore adopt the \lir\ from the fits to the Rieke et
al.\ models to all the IR data, which we report in Table~1.  If we
instead adopt the results from the fits to the Chary \& Elbaz models,
the star-formation efficiencies (SFEs) would decline, and gas
consumption  timescales would increase for the {$z=1.2-1.3$
  galaxies in our sample} studied here.   This would bring the SFEs further
in line with local spiral galaxies, strengthening that conclusion. 

\ifpreprint
\begin{figure*}
\centerline{
\includegraphics[width=0.5\textwidth]{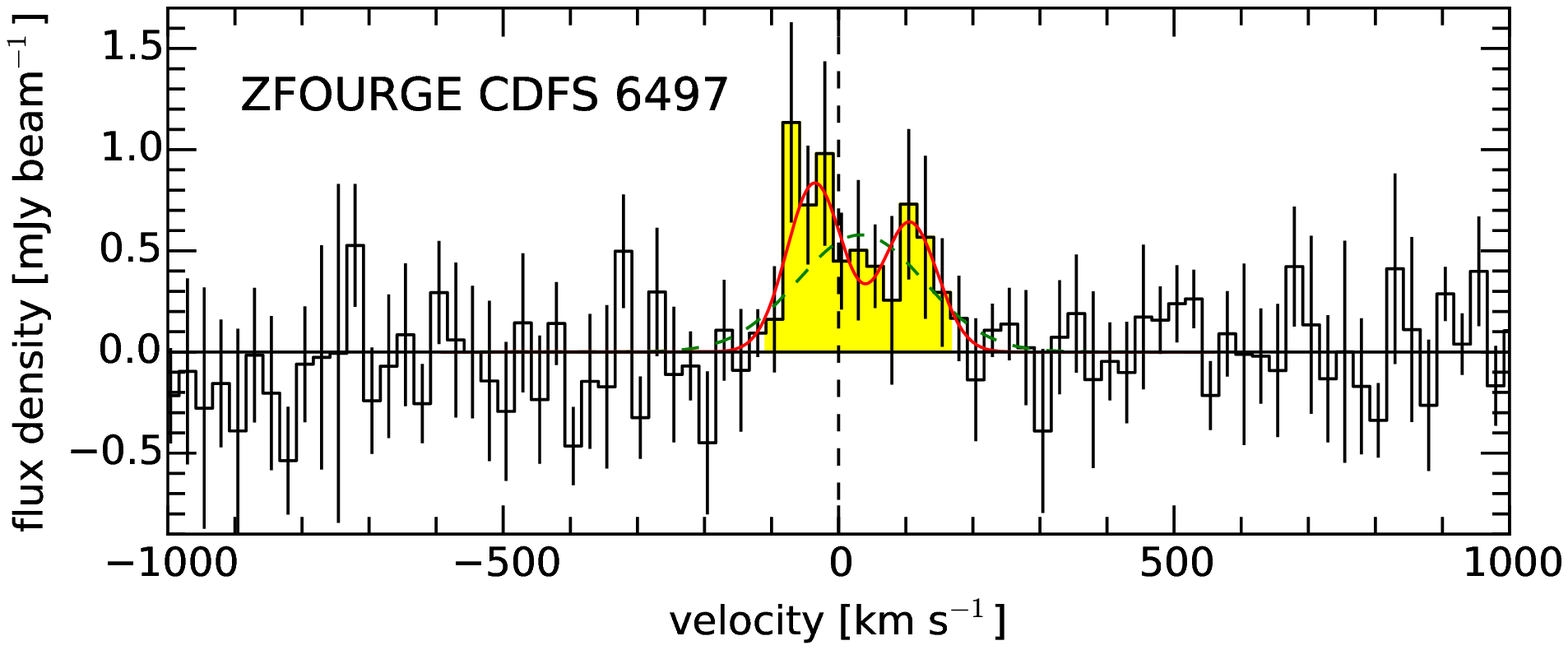}
\includegraphics[width=0.5\textwidth]{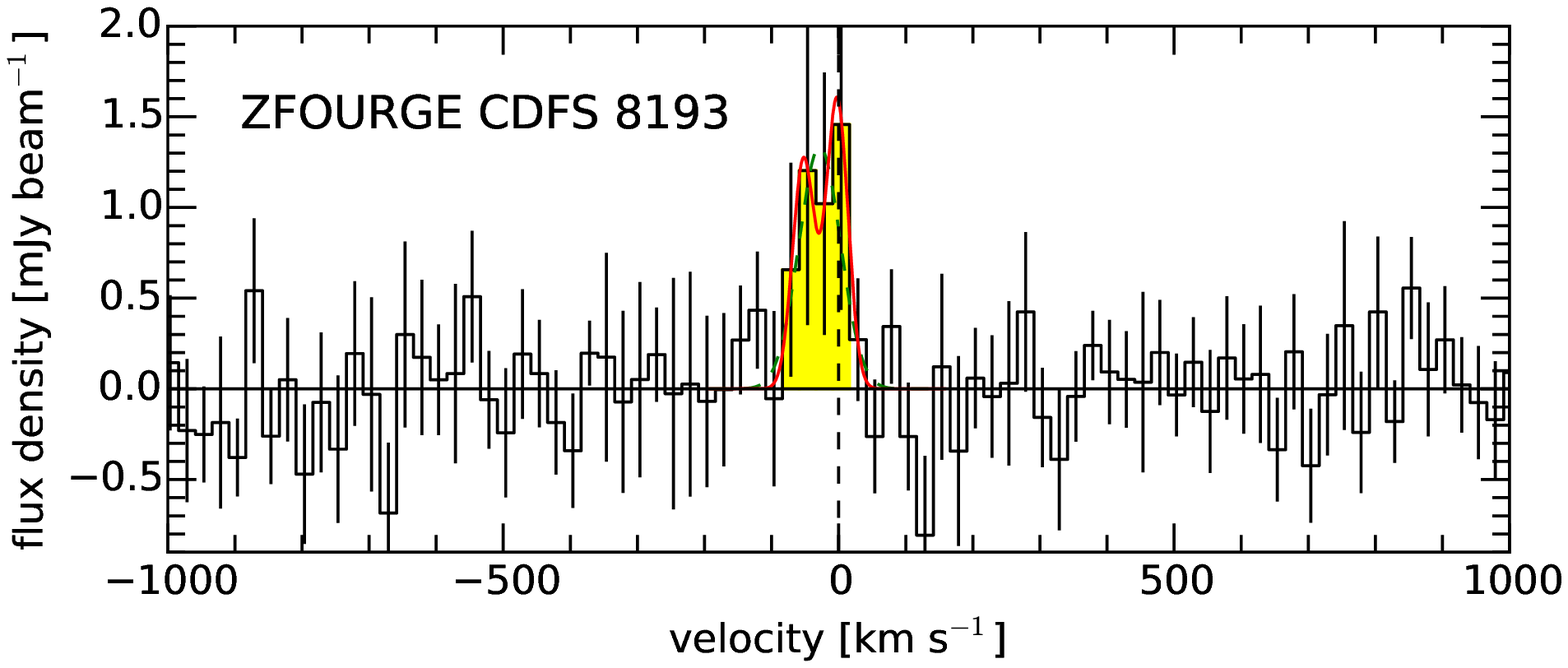}}
\centerline{
\includegraphics[width=0.5\textwidth]{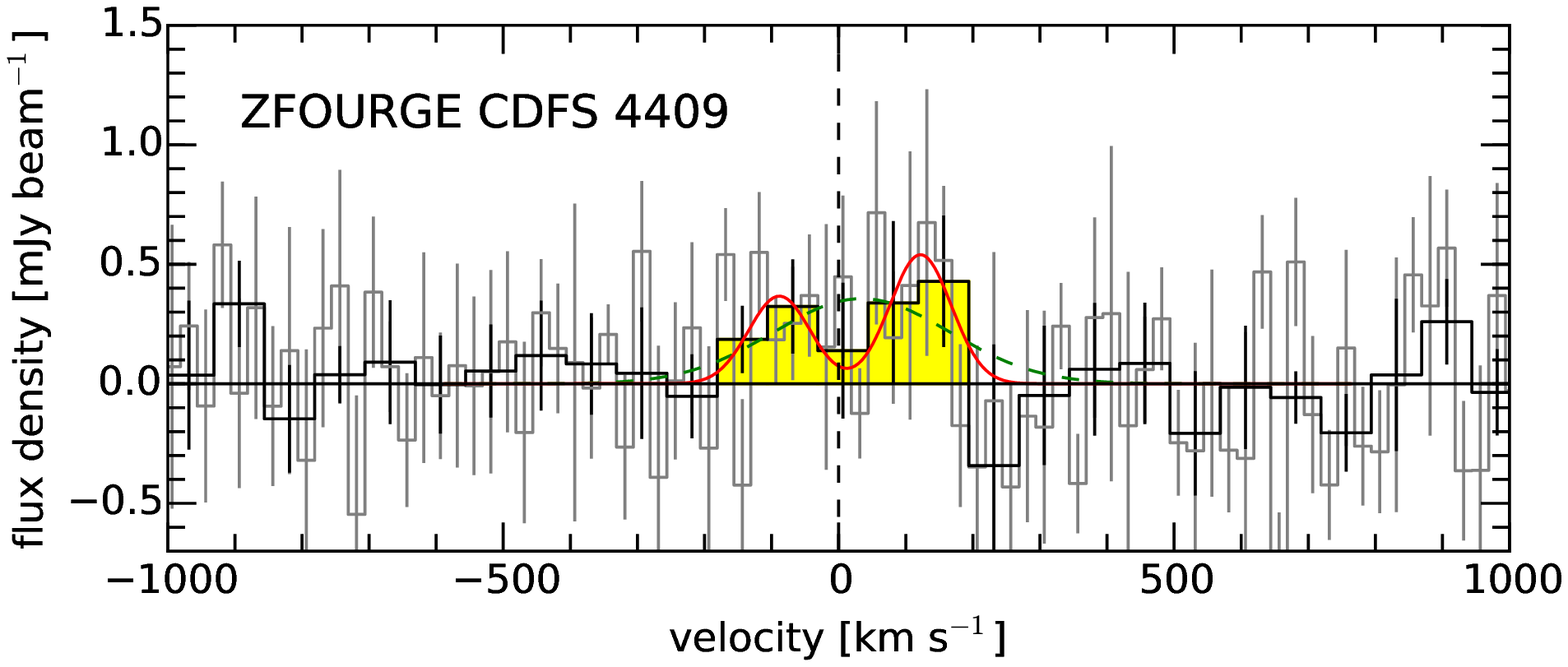}
\includegraphics[width=0.5\textwidth]{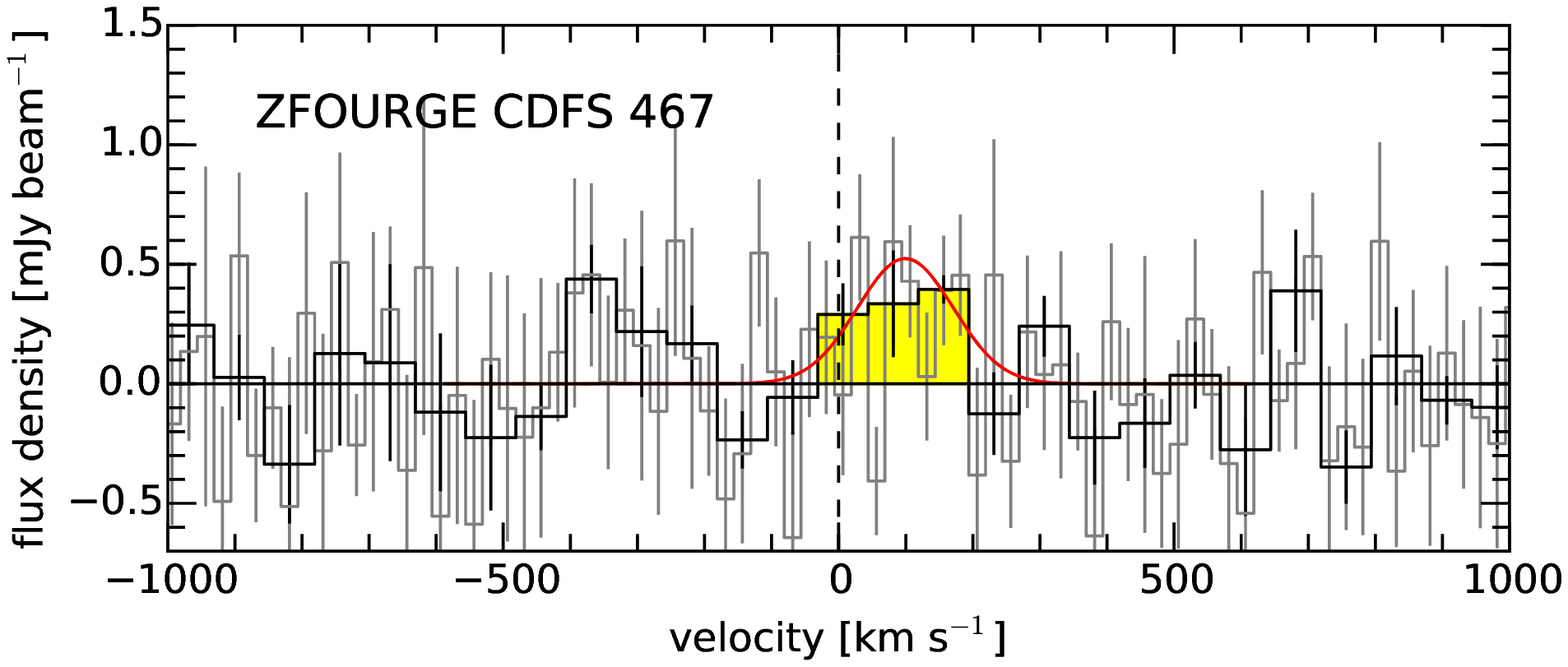}
}
\vspace{-2mm}
\caption{\textbf{Spectra of CO(3--2) for the {$\mathbf{log M_\ast/\msol}$$\mathbf{\simeq}$
    10.2 galaxies} at redshifts $\mathbf{z}$ = 1.2 to 1.3.}  In each panel, the spectra are
shown in 25 km s$^{-1}$ channels.  For objects ZFOURGE 467 and 4409,
the heavier-lined spectra are in 75 km s$^{-1}$ channels to improve
S/N.   The yellow-shaded regions indicate channels where postive
emission is detected at the expected location of the line.    In each
panel, the velocity is measured relative to the expected location of
the line from the optical spectroscopic redshift.     There are two
fits to each CO(3--2) line.    The green line shows a model with a
single Gaussian.  The red line shows a model with two
Gaussians. \label{fig:co_spec} }  
\vspace{-4mm}
\end{figure*}
 \fi

{The total IR luminosities for the $z$=1.2--1.3 galaxies in our
sample span $\lir$=$(1.5-3.2)$$\times$$10^{11}$~\lsol, as listed
in Table~1.  In these galaxies, most of the bolometric emission from
star  formation is emitted in the thermal IR.   In constrast, we
measure that the rest-frame UV (uncorrected for dust extinction)
contributes only 4--6\% to the total  SFR implied by the \lir\ in
these galaxies.  This is consistent with mean values measured in local
luminous IR galaxies.\cite{howe10}}

\section{ALMA Observations and Data Reduction}

Our Cycle 2 ALMA observations were taken between 2015 April 6 and 2015
May 2 in Band 4 with 36 antennas in the C34-2 configuration, which
provided a maximum baseline of 348.5 m.   For each source, we
configured ALMA to observe in four spectral windows, 1.875 GHz
per window, and spanning frequencies 134.48 to 156.90 GHz (depending
on the expected frequency of the CO[3--2] transition for each source).
We centered the CO(3--2) line in one of the spectral windows assuming
the optical spectroscopic redshift from the literature.\cite{vanz08}
The ALMA integrations ranged between 37.3 to 41.8 min on source.  One
source (ZFOURGE CDFS 6497) was erroneously observed twice, and received
double the exposure time.   The other spectral windows probe the
continuum of the line.   Flux, phase, and band-pass calibrators were
also obtained.  Supplementary Table~2 provides details about the observations for
each source. 

We reduced the data with CASA (Common Astronomy Software
Applications\cite{mcmull07}) version 4.5.0-REL with the calibration
script supplied by the National Radio Astronomy Observatory (NRAO).
We then ran the cleaning algorithm with natural weighting.  For the
spectral window containing the CO(3--2) line, we reduced the data with
channels of 25 km s$^{-1}$ and 75 km s$^{-1}$ with a cell size of
$0\farcs2$.  The angular sizes of the cleaned beam FWHM are given in
Table~S2, and this cell size gives 6--7 cells along the semi-minor
axis of the beam.

We also attempted to measure the continuum for each galaxy by cleaning
and combining the spectral windows excluding channels expected to have
CO emission.  We failed to detect any signal of the continuum; we also
therefore made no correction for the continuum to the CO line fluxes.

Supplementary figure~\ref{fig:co_spec} shows the spectra of the CO($J$=3--2)
emission for the four galaxies.  For each galaxy, there is positive emission in the
channels at the expected location of the CO(3--2) line.  To determine the peak of
emission we fit the spectra with models using  single- and
double-Gaussians.  The velocity offsets between the CO redshift and
redshift from the optical spectroscopy range from $-30$ to $+100$ km
s$^{-1}$. This is  consistent with uncertainties in the redshift
measurements.

We created total intensity maps of the CO(3--2) lines in each galaxy
by combining the channels showing positive emission around the
expected position of each line.   We also created first moment
(velocity) and second moment (velocity dispersion) maps to study
galaxy gas dynamics.        {From the primary-beam-corrected,
total intensity maps, we measured integrated flux densities for the
CO(3--2) transition, $I_{\mathrm{CO(3-2)}}$, for each galaxy in our
sample using the two-dimensional profile fitting tool in CASA.  These
are presented in Table~\ref{table:main}, and range from
$I_{\mathrm{CO(3-2)}}$=0.11--0.33 Jy km s$^{-1}$.   While the S/N of
the integrated values in Table~\ref{table:main} range from 2.3--7.7,
the detection significance (measured from the peak of the emission)
is much higher, where the S/N ranges from 4.8 to 13.7.}

{As discussed in the next section, the ALMA spectra in
Supplementary figure~\ref{fig:co_spec} show evidence for complex velocities, except
for ZFOURGE CDFS 467, where the data quality is lower.   The spectrum
of this object does show tentative, weak emission to the blue side of
the systemic redshift (at velocities $-$400 to $-$200 km s$^{-1}$).
However, the integrated emission from these channels is not
significant, having a S/N of $\simeq$2.0 at the peak.    When the
emission from the blue-side channels is summed with that on the red
side (where the object is detected), it lowers the overall significance
of the detection from 4.8 to 3.0.     We therefore do not include this
emission in the analysis of this object.  However, including it would
increase the CO luminosity by a factor of 1.6, making it more
consistent with the other objects in the sample.} 

\section{Velocity Shear in CO Emission} 

Our analysis of the spectra of the CO(3--2) emission in the four
{$z=1.2-1.3$ galaxies in our sample} in Supplementary figure~\ref{fig:co_spec}
shows that in many cases a double--Gaussian model fits better
reproduce the data than the single--Gaussian models.  This is
consistent with the expected signature of rotation.  Further evidence comes from observations of
velocity shear in the galaxies: some galaxies show spatial
variations in their velocity components.   For three of the galaxies
with the strongest emission, (ZFOURGE CDFS 4409, 6497, 8193) we measured
total CO(3--2) intensity maps separately from the channels blueward
(approaching) and redward (receding) relative to the velocity with the
minimum emission between the peaks.  Two of these galaxies (ZFOURGE
6497 and 8193) show velocity shear  (in the third object the
signal--to--noise is too low to centroid robustly the two separate
components).  Supplementary figure~\ref{fig:co_redblue} shows there are spatial
offsets in the centroids of the emission in the red and blue
components in these galaxies.  While the beam size precludes accurate
modelling of the velocity shear, the spatial variations are consistent
with rotation.  

\ifpreprint
\begin{figure*}[t]
\centerline{
\includegraphics[width=0.4\textwidth]{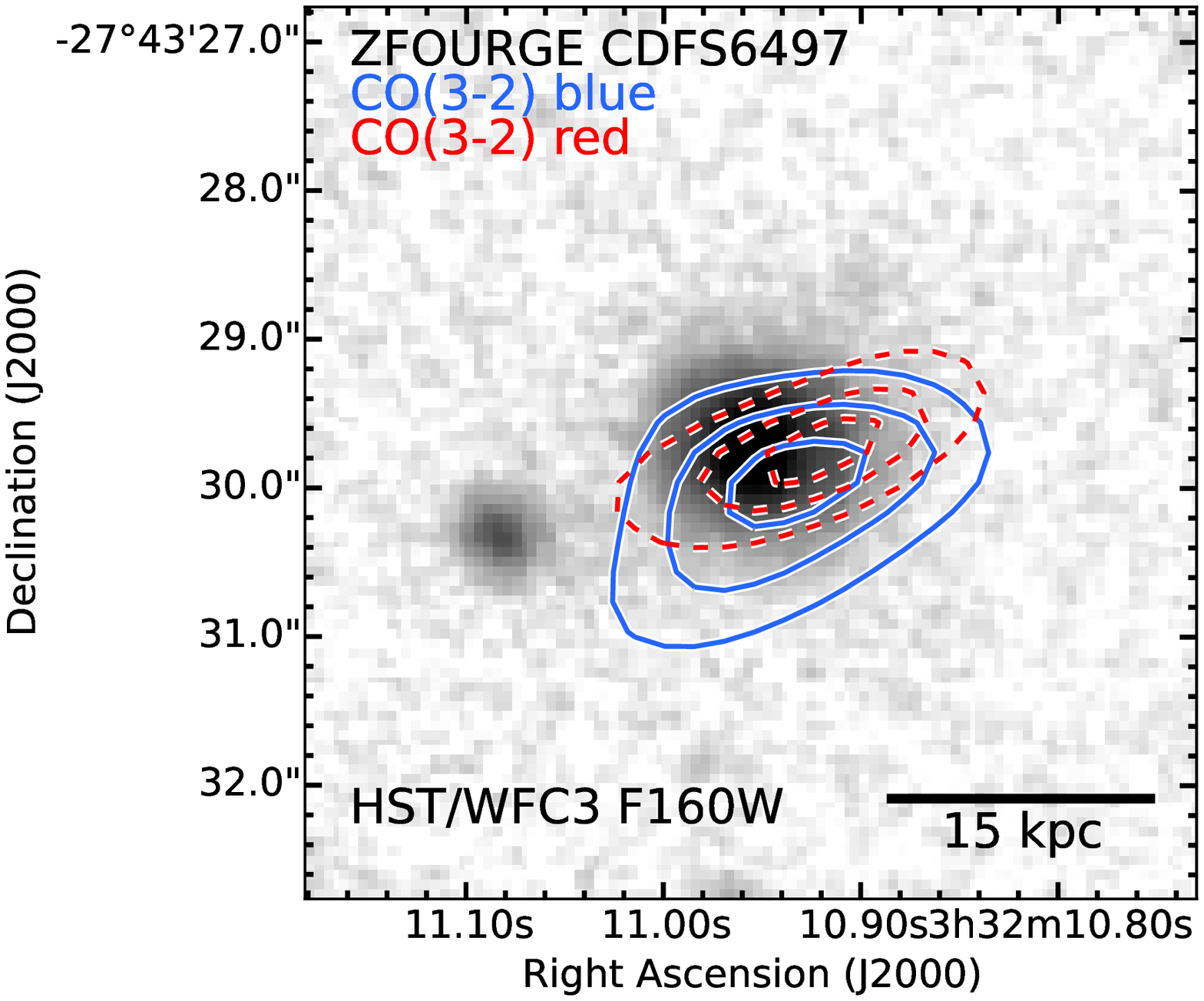}
\includegraphics[width=0.41\textwidth]{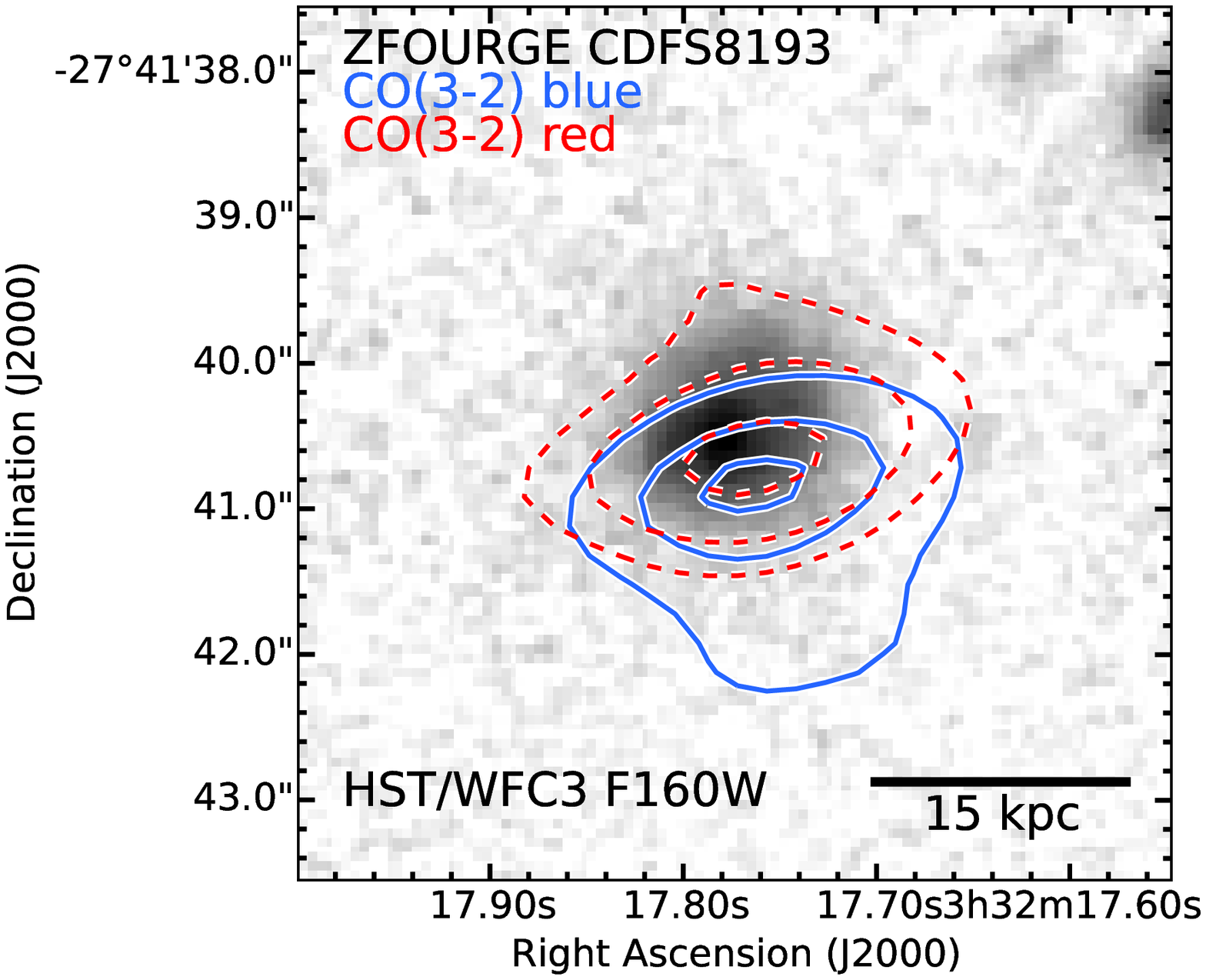}}
\vspace{-5mm}
\caption{\textbf{CO(3--2) maps of two galaxies with velocity shear:
spatial offsets in their velocity components}.  The contours show the
CO(3--2) emission from the blue-shifted (approaching) and red-shifted
(receding) emission.   The contours are overdrawn on the
HST/WFC3 F160W images from
CANDELS.\cite{grogin11,koek11}\label{fig:co_redblue}} \vspace{-0mm}
\end{figure*}
\fi

{Taken together, the CO spectra and spatial separation of the
approaching and receding velocity components provide reasonable
evidence for rotation in the $z=1.2-1.3$ {galaxies in our
sample}.   While the presence of double-peaked velocity structures in
the CO spectra could be expected for merging systems, we consider this
unlikely for two reasons.  First, the \textit{HST} morphologies of the
galaxies in our sample (see Figure~\ref{fig:co_hst}) show no
indications of double nuclei, which would be expected if a merger was
responsible for the CO velocity structure. Second, the galaxies in our
sample lie on the ``main-sequence'' of the SFR--$M_\ast$ relation
(Supplementary figure~\ref{fig:sfrms}), and they show no indications of
merger--induced starbursts on their star-formation efficiencies,
\lir/\lco\ (Figure~\ref{fig:lcolir}). } Direct confirmation of
rotation in these galaxies would require higher spatial resolution
kinematic data which is possible from ALMA in larger configurations,
but requires considerably more exposure time. 

 \section{Molecular Gas mass from CO emission}

The observed CO luminosity is proportional to the total cold molecular
gas mass, $M_\mathrm{gas}$.  At the temperatures and pressures of the
ISM in our galaxies, we expect that most of the gas exists in the
molecular phase,\cite{daddi10b} and therefore the molecular gas
accounts for the majority of baryons in the gas phase.  {The constant
of proportionality (the gas-mass--to--light ratio), $\alpha_\mathrm{CO}$, is given by}
\begin{equation}
\alpha_\mathrm{CO}  = M_\mathrm{gas} / \lco.
\end{equation}  
Based on the \lir/\lco\ ratios, the conditions in the
{$z$=1.2--1.3 galaxies in our sample }appear similar to normal
star-forming regions and star-forming disk galaxies, {which show values
of $\alpha_\mathrm{CO} \sim 4$ \msol\ (K km s$^{-1}$
pc$^2$)$^{-1}$,\cite{cari13} and traces the total amount of molecular
gas including a correction for helium.\cite{bola13}    The
CO--to--molecular gas conversion factor is $\alpha $=4.3 \msol\ (K km
s$^{-1}$ pc$^2$)$^{-1}$ for star-forming regions in the Galaxy and in
``normal'' star-forming galaxies.\cite{bola13}  The {galaxies
  in our sample} have  \lir/\lco\ ratios
consistent with other normal star-forming galaxies (see
figure~\ref{fig:lcolir}).   We therefore adopt $\alpha_\mathrm{CO} =
3.6$ \msol\ (K km s$^{-1}$ pc$^2$)$^{-1}$ found to apply to normal
star-forming (more massive) galaxies at these redshifts, and contains
a calibration uncertainty of 22\%.\cite{daddi10a} }   

{There is evidence that the CO--to--molecular gas ratio,
$\alpha_\mathrm{CO}$, varies with metallicity, $Z$, where the values
in the discussion above correspond to Solar values, $Z$=$Z_\odot$.
Theoretical work predicts that $\alpha_\mathrm{CO} \propto
Z^{-0.5}$,\cite{feld12,nara12} while empirical measurements at
redshifts $z > 1$ suggest a possible steeper relation,
$\alpha_\mathrm{CO} \propto Z^{-1.2}$ to $Z^{-1.8}$.\cite{genzel12} If
the galaxies in our ALMA sample have $Z$$<$$Z_\odot$, one may expect
an increase in $\alpha_\mathrm{CO}$.  Work on the
stellar-mass---metallicity ($M^\ast$--$Z$) relation at $0.9 < z < 1.3$
shows that star-forming galaxies in the mass range of our sample
should have metallicities between
0.6--1.0~$Z_\odot$.\cite{perez09,zahid11,stott13}  This implies a
higher $\alpha_\mathrm{CO}$ (and higher gas masses) by at most a
factor of $\sim$2.  This would correspond to even more dramatic
evolution in the gas fraction from $z\sim 1.3$ to the present for
Milky-way mass galaxies.   For this reason, we adopt the (more
conservative) CO--to-molecular gas ratio for Solar metallicity,
$\alpha_\mathrm{CO}=3.6$ \msol (K km s$^{-1}$ pc$^{2}$)$^{-1}$, as
stated in the discussion above.} 

\section{Conventions}

Throughout, we assume a Chabrier initial mass function\cite{chab03}
when deriving stellar masses and SFRs.  For all cosmological
calculations, we assume $\Omega_m = 0.3$, $\Omega_\Lambda = 0.7$, and
$H_0 = 70$~km s$^{-1}$ Mpc$^{-1}$, consistent with the recent
constraints from Planck\cite{planck15} and the local distance
scale.\cite{riess16}

\ifpreprint \else 
%
%
%
%
\pagebreak
\tableOne

\fi

\end{document}